\pdfoutput=1
\documentclass[10pt,a4paper]{article}
\usepackage{jheppub,amsmath,color,graphicx,hyperref,slashed,multirow,lmodern,upquote,titlesec}
\usepackage[utf8]{inputenc}
\usepackage[T1]{fontenc}
\usepackage[onehalfspacing]{setspace}

\newcommand{\alphas}{\alpha_{\text{s}}}
\newcommand{\lqp}{{\text{LQ}\,\text{LQ}^*}}
\newcommand{\mlq}{{m_{\text{LQ}}}}
\newcommand{\pwb}{{\textsc{POWHEG-BOX}}}
\newcommand{\fa}{{\textsc{FeynArts}}}
\newcommand{\fc}{{\textsc{FormCalc}}}
\newcommand{\collier}{{\textsc{Collier}}}
\newcommand{\mgf}{{\textsc{MadGraph~4}}}
\newcommand{\fr}{{\textsc{FeynRules}}}
\newcommand{\mgamc}{{\textsc{MadGraph5\_aMC@NLO}}}
\newcommand{\py}{{\textsc{Pythia~8}}}
\newcommand{\MSbar}{{\overline{\text{MS}}}}
\def\sss{\scriptscriptstyle}
\newcommand{\be}{\begin{equation}}
\newcommand{\ee}{\end{equation}}
\def\bsp#1\esp{\begin{split}#1\end{split}}
\def\bpm{\begin{pmatrix}}
\def\epm{\end{pmatrix}}

\def\lag{{\cal L}}
\def\dR{d_{\sss R}}
\def\dRbar{{\bar d}_{\sss R}}
\def\eR{\ell_{\sss R}}
\def\eRbar{{\bar e}_{\sss R}}
\def\Ll{L_{\sss L}}

\def\QL{Q_{\sss L}}
\def\QLbar{\bar Q_{\sss L}}
\def\uR{u_{\sss R}}
\def\uRbar{{\bar u}_{\sss R}}

\titleformat{\paragraph}
{\normalfont\normalsize\bfseries}{\theparagraph}{1em}{}
\titlespacing*{\paragraph}
{0pt}{3.25ex plus 1ex minus .2ex}{1.5ex plus .2ex}

\title{Scalar leptoquark pair production at the LHC: precision predictions in the era of flavour anomalies}

\author[a]{Christoph~Borschensky}
\author[b,c]{\!\!, Benjamin~Fuks}
\author[d]{\!\!, Anna~Kulesza}
\author[d]{\! and Daniel~Schwartl\"ander}

\affiliation[a]{Institute for Theoretical Physics, University of T\"ubingen,
  Auf der Morgenstelle 14, 72076 T\"ubingen, Germany}
\affiliation[b]{Laboratoire de Physique Th\'eorique et Hautes Energies (LPTHE), UMR 7589, Sorbonne Universit\'e et
CNRS, 4 place Jussieu, 75252 Paris Cedex 05, France}
\affiliation[c]{Institut Universitaire de France, 103 boulevard Saint-Michel,
  75005 Paris, France}
\affiliation[d]{Institute for Theoretical Physics, WWU M\"unster, D-48149 M\"unster,
  Germany}

\emailAdd{christoph.borschensky@uni-tuebingen.de}
\emailAdd{fuks@lpthe.jussieu.fr}
\emailAdd{anna.kulesza@uni-muenster.de}
\emailAdd{d\_schw20@uni-muenster.de}

\abstract{
We comprehensively examine precision predictions for scalar leptoquark pair production at the LHC. In particular, we investigate the impact of lepton $t$-channel exchange diagrams that are potentially relevant in the context of leptoquark scenarios providing an explanation for the flavour anomalies. We also evaluate the corresponding total rates at the next-to-leading order in QCD. Moreover, we complement this calculation with the resummation of soft-gluon radiation at the next-to-next-to-leading logarithmic accuracy, hence providing the most precise predictions for leptoquark pair production at the LHC to date. Relying on a variety of benchmark scenarios favoured by the anomalies, our results exhibit an interesting interplay between the $t$-channel diagram contributions, the flavour texture satisfied by the leptoquark Yukawa couplings, the leptoquark masses and their representations under the Standard Model gauge group, as well as the chosen set of parton densities used for the numerical evaluations. The net effect on a cross section turns out to be very non-generic and ranges up to about 60\% with respect to the usual next-to-leading-order predictions in QCD ({\it i.e.}\ without any $t$-channel contribution) for some scenarios considered. Dedicated calculations are thus required for any individual leptoquark model that could be considered in a collider analysis in order to assess the size of the studied corrections. In order to facilitate such calculations we provide dedicated public numerical packages.
}

\begin{document}


\maketitle
\flushbottom

\section{Introduction}
Scalar leptoquarks are hypothetical bosonic particles beyond the Standard Model of particle physics that carry both lepton and baryon numbers. They therefore generically couple simultaneously to one quark and one lepton. Whereas leptoquarks have been initially proposed in the context of Grand Unification~\cite{Pati:1973uk,Pati:1974yy,Georgi:1974sy,Fritzsch:1974nn,Senjanovic:1982ex,Frampton:1989fu,Murayama:1991ah}, they also arise in many extensions of the Standard Model. These include, for example, technicolour and composite models~\cite{Dimopoulos:1979es,Eichten:1979ah,Farhi:1980xs,Schrempp:1984nj,Lane:1991qh}, low-energy manifestations of superstring models~\cite{Hewett:1988xc}, as well as $R$-parity-violating supersymmetric scenarios~\cite{Farrar:1978xj,Barbier:2004ez}. Leptoquarks generally possess a fractional electric charge and lie in the fundamental (or anti-fundamental) representation of the QCD gauge group $SU(3)_C$. All other properties, such as their weak isospin quantum numbers, depend on the considered model of new physics.

Scalar leptoquarks have attracted a lot of attention over the recent years due to the so-called flavour anomalies~\cite{Lees:2012xj,Abdesselam:2019wac,Abdesselam:2019lab,Belle:2019rba,Aaij:2017vbb,Aaij:2017uff,Aaij:2019wad,Aaij:2021vac} and the long-standing theory-experiment discrepancy related to the anomalous magnetic moment of the muon~\cite{Bennett:2006fi,Abi:2021gix,Aoyama:2020ynm}. Leptoquarks can indeed mediate violations of lepton flavour universality, enabling a reduction of the discrepancies between the predictions of the Standard Model and the observations in the context of several flavour observables, or even in some cases restoring agreement between theory and experiment. For this to happen, the Yukawa couplings of the scalar leptoquark to a quark and a lepton must usually be large. Interestingly, scalar leptoquark solutions to the above anomalies still persist after the recent LHCb update, as shown for instance in refs.~\cite{Crivellin:2020tsz,Angelescu:2021lln,Nomura:2021oeu,Marzocca:2021azj,FileviezPerez:2021xfw,Murgui:2021bdy,Singirala:2021gok}.

Consequently, evidence for leptoquark production and decay are widely searched for at the LHC, in a variety of channels dedicated each to a specific signature. No signal has been seen so far and new limits have been set on phenomenologically-viable leptoquark models. The analysis of the LHC run~2 dataset constrains third generation leptoquarks to be heavier than about 1.0--1.8~TeV~\cite{Aaboud:2019bye,Aad:2020sgw,ATLAS:2021oiz,Aad:2021jmg,Sirunyan:2018jdk,Sirunyan:2018vhk,CMS:2020wzx}, the exact bounds depending on the benchmark scenario and the considered leptoquark decay pattern. It moreover also constrains first and second generation leptoquarks to be heavier than 1.2--1.8~TeV~\cite{Aaboud:2019jcc,Aad:2020iuy,Sirunyan:2018btu,Sirunyan:2018ryt}. In addition, the ATLAS collaboration has also carried out a search for the production of a pair of leptoquarks that couple inter-generationally, as favoured by the $B$-anomalies, which has led to limits of about 1.5~TeV~\cite{Aad:2020jmj}. Those bounds can however be reduced as soon as several leptoquark decay modes exist.

\begin{figure}[t]
	\centering
	\begin{tabular}{ccc}
		\includegraphics[width=.3\textwidth]{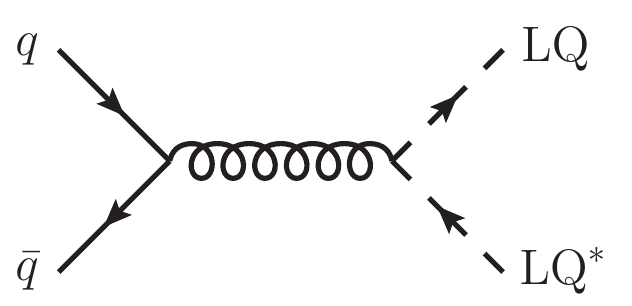} & \includegraphics[width=.3\textwidth]{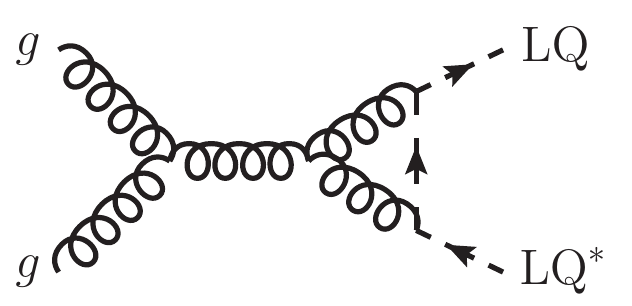} & \includegraphics[width=.3\textwidth]{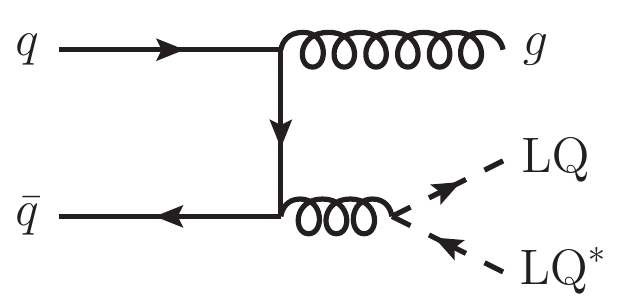}\\
		(a) & (b) & (c)\\[2mm]
		\includegraphics[width=.3\textwidth]{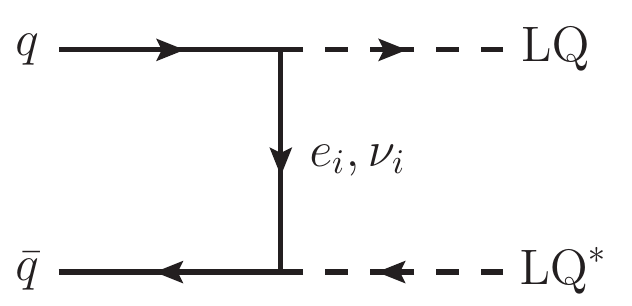} & \includegraphics[width=.3\textwidth]{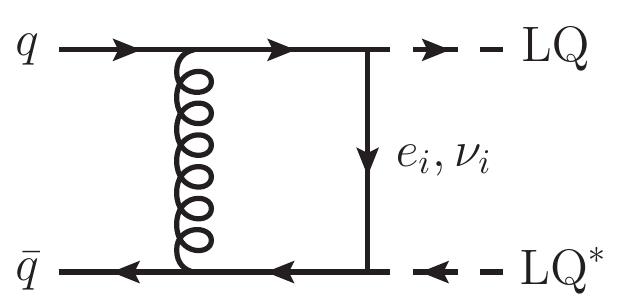} & \includegraphics[width=.3\textwidth]{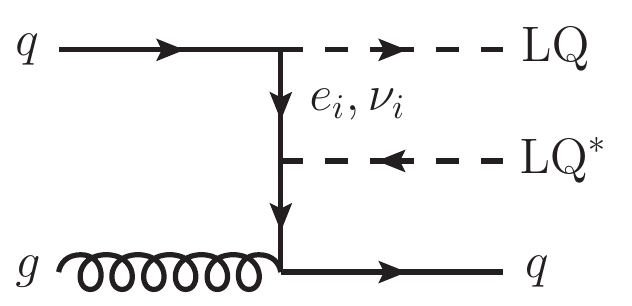}\\
		(d) & (e) & (f)
	\end{tabular}
	\caption{Selection of Born, virtual, and real emission Feynman diagrams for partonic processes contributing to leptoquark pair production at the LHC. Upper row: pure-QCD contributions; lower row: leptonic $t$-channel contributions.}
	\label{th:feynmandiagrams}
\end{figure}

All the above-mentioned searches exploit leptoquark pair production, and some of them additionally consider single leptoquark production. In the context of pair production, the lepton-leptoquark-quark Yukawa couplings are always assumed small, so that the pair-production mechanism can be approximated by its pure QCD contributions (as illustrated by the representative Feynman diagrams (a)--(c) of figure~\ref{th:feynmandiagrams}). Any $t$-channel leptonic exchange contributions like the ones represented by the diagrams (d)--(f) of figure~\ref{th:feynmandiagrams} are thus omitted. The signal simulation strategy that is then followed by the ATLAS and CMS collaborations, and that is at the heart of the limit setting procedure, differs. CMS signal simulations rely on leading-order (LO) calculations matched with parton showers and are handled either with \mgamc~\cite{Alwall:2014hca} and the \fr/UFO~\cite{Alloul:2013bka,Christensen:2009jx,Degrande:2011ua,Degrande:2014vpa} model developed in ref.~\cite{Dorsner:2018ynv}, or with \py~\cite{Sjostrand:2014zea}. The generated LO signals are then re-weighted so that next-to-leading-order (NLO) QCD $K$-factors for the total rate are included~\cite{Kramer:1997hh,Kramer:2004df}. ATLAS simulations rely in contrast on NLO-QCD accurate calculations matched with parton showers, using the \fr/UFO model developed in ref.~\cite{Mandal:2015lca}. Signal rates are then re-scaled to incorporate approximate next-to-next-to-leading-order (NNLO) corrections and the resummation of the threshold logarithms at the next-to-next-to-leading-logarithmic (NNLL) accuracy, as obtained from the analogous results for stop pair production~\cite{Beenakker:2016gmf}.

In the light of supporting a leptoquark explanation for the flavour anomalies and for the discrepancies inherent to the anomalous magnetic moment of the muon, the lepton-leptoquark-quark Yukawa couplings cannot be neglected and have to be of ${\cal O}(1)$. Additionally, the leptoquark mass should lie in the TeV-regime. Leptonic $t$-channel exchange contributions to leptoquark pair production could therefore potentially play an important role\footnote{Notably, a recent study~\cite{Dorsner:2021chv} has pointed out that the $t$-channel exchange mechanism also enables novel off-diagonal production modes which have not been considered experimentally yet.}, and  the resummation of the threshold logarithms could significantly affect the cross section. We have recently shown that those two components of the most advanced theoretical calculations for scalar leptoquark pair production interplay with parton density effects and leptoquark flavour decompositions in a non-trivial and very model-dependent manner~\cite{Borschensky:2020hot}. In the present work, we comprehensively study these effects in different leptoquark scenarios that are compatible with an explanation for the flavour anomalies and that are consistent with current experimental constraints. We show predictions for total cross sections at the NLO+NNLL precision. Our predictions consistently include $t$-channel exchange contributions.

In the rest of this work, in section~\ref{sec:theory} we begin with a description of our theoretical framework. We then briefly review leptoquark pair-production cross sections at fixed order, and the resummation of large threshold logarithmic corrections. In section~\ref{sec:numerics}, we detail our implementations in the \mgamc~\cite{Alwall:2014hca} and \pwb~\cite{Nason:2004rx,Frixione:2007vw,Alioli:2010xd} frameworks, that are the key ingredients of our numerical code. We then dedicate section~\ref{sec:results} to our results, illustrating the impact of the parton densities, the leptonic $t$-channel exchange contributions, the flavour structure of the leptoquark and the resummation of the threshold logarithms on the production cross sections. We summarise our findings in section~\ref{sec:conclusions}.

\section{Theory}\label{sec:theory}
In order to set up a generic framework allowing for the most advanced calculations of scalar leptoquark pair production in QCD, we consider a simplified model in which the Standard Model (SM) is minimally extended. This model is briefly detailed in section~\ref{sec:model}, in which we also fix our notation and conventions. We next introduce in section~\ref{sec:benchmarks} a set of benchmark scenarios that will serve as a basis for our predictions. We consider both simple models that will be useful to understand specific effects of our calculations, as well as scenarios that are relevant in the light of the flavour anomalies. Details on leptoquark pair production at hadron colliders are then provided in section~\ref{sec:lqprod}, both for what concerns fixed-order calculations and soft-gluon resummation.

\subsection{A simplified and general model for scalar leptoquark pair production at the LHC}
\subsubsection{Lagrangian and models}\label{sec:model}
We consider a simplified model in which we supplement the Standard Model by all possible five species of scalar leptoquarks which couple to the SM fermions, {\it i.e.}\ $S_1$, $\tilde S_1$, $R_2$, $\tilde R_2$ and $S_3$. Following standard notation~\cite{Buchmuller:1986zs,Dorsner:2016wpm}, those leptoquarks lie in the $({\bf 3},{\bf 1})_{\sss -1/3}$, $({\bf 3},{\bf 1})_{\sss -4/3}$, $({\bf 3},{\bf 2})_{\sss 7/6}$, $({\bf 3},{\bf 2})_{\sss 1/6}$ and $({\bf 3},{\bf 3})_{\sss-1/3}$ representation of the SM gauge group. The corresponding electroweak multiplets can then be written in terms of their component fields as
\be\renewcommand{\arraystretch}{1.4}\bsp
  & S_1        = S_1^{(-1/3)} \ , \qquad
    \tilde S_1 = \tilde S_1^{(-4/3)} \ , \qquad
    R_2        = \bpm R_2^{(+5/3)} \\ R_2^{(+2/3)}\epm\ ,\qquad
    \tilde R_2 = \bpm \tilde R_2^{(+2/3)} \\ \tilde R_2^{(-1/3)}\epm\ ,\\
  & \hspace{4.5cm}S_3 = \bpm \frac{1}{\sqrt{2}}S_3^{(-1/3)} & S_3^{(+2/3)}\\
              S_3^{(-4/3)} & -\frac{1}{\sqrt{2}}S_3^{(-1/3)} \epm,
\esp\ee
where the superscripts refer to the electric charge of the various fields (and the subscripts to their respective $SU(2)_L$ representation). In the above expressions, we have used a matrix representation for the electroweak triplet that is defined as
\be
  (S_3)^i{}_j = \frac{1}{\sqrt{2}} (\sigma_k)^i{}_j S_3^k,
\ee
where $S_3^k$ carries an $SU(2)_L$ adjoint index ($k=1,2,3$), $(S_3)^i{}_j$ carries one fundamental ($i=1,2$) and one antifundamental ($j=1,2$) index, and $\sigma_k$ are the usual Pauli matrices. The leptoquark interaction, kinetic and mass terms are described by the Lagrangian
\be\bsp
   \lag_{\rm LQ} = &\ \lag_{\rm kin.} +
   {\bf y_{\sss 1}^{\sss RR}} \uRbar^c \eR S_1^\dagger
  +{\bf y_{\sss 1}^{\sss LL}} \big(\QLbar^c \!\cdot\! \Ll\big) S_1^\dagger
  +{\bf \tilde{y}_{\sss 1}^{\sss RR}} \dRbar^c \eR \tilde S_1^\dagger
  +{\bf y_{\sss 2}^{\sss LR}} \eRbar \QL R_2^\dag
 \\ &\ +
   {\bf y_{\sss 2}^{\sss RL}} \uRbar\big(\Ll\!\cdot\!R_2\big)
  +{\bf \tilde{y}_{\sss 2}^{\sss RL}} \dRbar\big(\Ll\!\cdot\!\tilde R_2\big)
  +{\bf y_{\sss 3}^{\sss LL}} \big(\QLbar^c\!\cdot\!\sigma_k \Ll\big) \big(S_3^k\big)^\dag
  + {\rm h.c.},
\esp\label{eq:lag}\ee
in which we only consider leptoquark Yukawa couplings involving one Standard Model lepton (the right-handed neutrinos being thus excluded) and one Standard Model quark. Moreover, the leptoquark gauge-invariant kinetic and mass terms are collected into the Lagrangian $\lag_{\rm kin.}$. All flavour indices are suppressed for clarity, the ${\bf y/\tilde{y}}$ couplings being $3\times 3$ matrices in the flavour space. The first index of any $y_{ij}$ element of these matrices refers to the quark generation and the second index to the lepton generation. Additionally, the dot product appearing in~\eqref{eq:lag} represents the invariant product of two fields lying in the (anti)fundamental representation of $SU(2)_L$. In our notation, the $\QL$ and $\Ll$ spinors are the weak doublets of Standard Model left-handed quarks and leptons, and the $\uR$, $\dR$ and $\eR$ spinors are the corresponding weak singlets.

\subsubsection{Benchmark scenarios}\label{sec:benchmarks}
\paragraph{Simple scenarios}\label{sec:benchmsimple}
In order to be able to analyse the impact of different improvements to scalar leptoquark pair production that we examine in this study, we first consider two simple setups in which the Standard Model is extended by exactly one species of leptoquarks. Whilst strictly speaking these scenarios cannot explain the entire set of flavour anomalies, they represent interesting benchmarks allowing us to understand all the distinct features that enter our calculations. Therefore, in a first step we make use of them for this purpose, and we consider more relevant scenarios in the light of the flavour anomalies in the next step.

We focus on models that feature either an $S_1^{(-1/3)}$ leptoquark eigenstate or an $R_2^{(+5/3)}$ eigenstate. The leptoquark mass is chosen to vary in the 1--2~TeV range, and we enforce that the leptoquark-quark-lepton Yukawa couplings follow a specific pattern. The models are defined by
\be\bsp
 S_1:&\ \mlq\in[1, 2]~{\rm TeV}, \qquad y_{1,22}^{LL} = -0.15,\qquad y_{1,32}^{LL} = 3;\\
 R_2:&\ \mlq\in[1, 2]~{\rm TeV}, \qquad y_{2,22}^{RL} = 1.5,
\esp\ee
with all other entries of the Yukawa coupling matrices being zero. As already mentioned above, while these two classes of simple scenarios are motivated by the flavour anomalies, they are not sufficient to explain them all.

For the $S_1$ case, we have fixed $y_{1,22}^{LL} = -0.15$ and $y_{1,32}^{LL} = 3$. For leptoquark masses in the 1--2~TeV range, those parameter space configurations allow us to accommodate the $R_{K^{(*)}}$ anomalies, where
\be
  R_{K^{(*)}} = \frac{{\rm BR}(B\to K^{(*)}\mu^+\mu^-)}{{\rm BR}(B\to K^{(*)}e^+e^-)},
\label{eq:RK}\ee
but not the $R_{D^{(*)}}$ ones, where
\be
  R_{D^{(*)}} = \frac{{\rm BR}(B\to D^{(*)}\tau\bar\nu)}{{\rm BR}(B\to D^{(*)}\ell\bar\nu_\ell)},
\label{eq:RD}\ee
for $\ell=e, \mu$~\cite{Angelescu:2018tyl}. The large adopted $y_{1,k2}^{LL}$ values (with $k=2, 3$) indeed yield a decrease of the size of the numerator appearing in~\eqref{eq:RK}, reducing hence the disagreement with data. In order to additionally provide an explanation for the $R_{D^{(*)}}$ anomalies, one option would be to turn on the $y_1^{LL}$ and $y_1^{RR}$ couplings of the $S_1$ leptoquark to tau leptons and second-generation quarks, thus changing the numerator of eq.~\eqref{eq:RD}. It has however been shown that this could only lead to a moderate reduction of the existing tensions between predictions and data, provided that the couplings are not too large.

In the $R_2$ case, we have only turned on the $y_{2,22}^{RL}$ coupling that we have set to 1.5. In such a scenario, an explanation for the $R_{K^{(*)}}$ anomalies is once again provided through a decrease of the numerator in eq.~\eqref{eq:RK}, although the agreement with data is only restored at most at the $2\sigma$ level for reasonable values of the Yukawa couplings~\cite{Angelescu:2018tyl}. Such a large Yukawa coupling is however in tension with existing LHC limits, and this coupling texture does not lead to any explanation for the $R_{D^{(*)}}$ anomalies. Once the $R_2$ leptoquark couples to muons, there is indeed no simultaneous explanation for all flavour anomalies that is viable from the standpoint of other constraints. Despite this situation, leptoquark scenarios including an $R_2$ state as a solution to the flavour anomalies have however been recently resurrected by considering not too large complex-valued couplings and by including a connection with the electron instead of the muon~\cite{Popov:2019tyc}. This class of scenarios are considered in the next subsection.

\paragraph{Phenomenologically-viable $R_2$ models}\label{sec:benchmr2}
\renewcommand{\arraystretch}{1.5}\setlength\tabcolsep{10pt}
\begin{table}
 \centering
 \begin{tabular}{l | c | ccc}
   & $y^{RL}_{2,23}$ & $y^{LR}_{2,33}$ & $y^{LR}_{2,21}$ & $y^{LR}_{2,31}$
  \tabularnewline
  \hline
  $a_1$ & $1.84+1.84i$ & $0.354+0.354i$ & $-0.015i$ & $0.262+0.262i$
  \tabularnewline
  $a_2$ & $0.309+0.951i$ & $0.951+0.309i$ & $0.011-0.011i$ & $0.37i$
  \tabularnewline
 \end{tabular}
 \caption{Single $R_2$ leptoquark scenarios providing an explanation for the flavour anomalies according to the fit of ref.~\cite{Popov:2019tyc}. The two benchmark points have been selected from the $1\sigma$ ranges of model parameters allowing to explain the anomalies. All other elements of the Yukawa coupling matrices are equal to zero, and the leptoquark mass is set in both cases to $\mlq = 1$\,TeV.}
 \label{tab:benchmarksR2}
\end{table}

Recently, a new promising route to solve all the flavour anomalies with a unique species of scalar leptoquarks has been proposed~\cite{Popov:2019tyc}. Its core idea consists of an extension of the Standard Model featuring a single $R_2$ state that couples to taus (to address the $R_{D^{(*)}}$ anomalies), and to electrons (to address the $R_{K^{(*)}}$ anomalies). By tuning the corresponding Yukawa couplings, it becomes possible to act on the denominator in the $R_{K^{(*)}}$ ratio, instead of acting on its numerator as when we consider that the $R_2$ leptoquark couples to muons. More precisely, such a solution enables us to recover an agreement between data and theory while keeping all new physics couplings moderately large, in contrast to the requirement that some Yukawa couplings should be of about 2 or 3 when leptoquark couplings to muons are considered.

We study scenarios in which the $R_2$ leptoquark simultaneously couples to electrons and taus. On the basis of the fit to data presented in ref.~\cite{Popov:2019tyc}, we consider two benchmark points $a_1$ and $a_2$ with $\mlq = 1~{\rm TeV}$, and for which all non-vanishing entries of the leptoquark Yukawa matrices are given in table~\ref{tab:benchmarksR2}. Whereas these setups result in mild tensions between predictions and measurements of the BR$(B_c\to\tau\nu)$ branching ratio, they can in principle be alleviated through non-minimal scenarios featuring both an $R_2$ and an $S_3$ leptoquark. Such a two-leptoquark configuration, additionally motivated by radiative neutrino mass models, is considered in the next subsection in a Grand Unified Theory context (and with a different Yukawa coupling texture).

\paragraph{A two-leptoquark model inspired by Grand Unification: $R_2$ and $S_3$}\label{sec:benchmr2s3}
\renewcommand{\arraystretch}{1.5}\setlength\tabcolsep{8pt}
\begin{table}
 \centering
 \begin{tabular}{l | c | cc | cccc }
  & $y_{2,33}^{LR}$ & $y_{2,22}^{RL}$ & $y_{2,23}^{RL}$ & $y_{3,22}^{LL}$ & $y_{3,23}^{LL}$ & $y_{3,32}^{LL}$ & $y_{3,33}^{LL}$
  \tabularnewline
  \hline
  $b_1$ & $-0.18734+1.12287i$ & $0.265001$ & $1.17382$ & $-0.010$ & $-0.045$ & $-0.265$ & $-1.173$
  \tabularnewline
  $b_2$ & $-0.18734+1.12287i$ & $0.37353$ & $1.59511$ & $-0.014$ & $-0.061$ & $-0.373$ & $-1.594$
  \tabularnewline
 \end{tabular}
 \caption{Two-leptoquark explanations to the flavour anomalies when the Standard Model is extended with an $R_2$ and an $S_3$ leptoquark. We consider scenarios originating from the fit of ref.~\cite{Becirevic:2018afm} and its more recent update~\cite{Becirevic:2020pc}. Benchmark point $b_1$ corresponds to the best fit value and $b_2$ is chosen inside the $2\sigma$ region of the fit. All other elements of the Yukawa coupling matrices are equal to zero, and the leptoquark masses are $m_{R_2} = 1.3$\,TeV and $m_{S_3} = 2$\,TeV.}
 \label{tab:benchmarksR2plusS3}
\end{table}

Models featuring several leptoquark species at TeV-scale energies that can explain both the $R_{K^{(*)}}$ and $R_{D^{(*)}}$ anomalies also emerge from Grand Unified Theories in which the Standard Model is embedded into an $SU(5)$ gauge symmetry~\cite{Becirevic:2018afm}. In this context, the Standard Model fermions appear as components of fields lying in the $\overline{{\bf 5}}$ and ${\bf 10}$ representations of $SU(5)$, and the $R_2$ and $S_3$ leptoquarks are admixtures of the components of scalar fields lying in the ${\bf 45}$ and ${\bf 50}$ representations of the Grand Unified gauge group respectively. After breaking the $SU(5)$ symmetry, two leptoquark mass eigenstates with the respective quantum numbers of the $R_2$ and $S_3$ states can lie in the TeV regime, all other leptoquark states being decoupled. The interactions of these light leptoquarks with the Standard Model fermions are then given as the $R_2$ and $S_3$ interactions appearing in the Lagrangian~\eqref{eq:lag}.

By enforcing that the two leptoquarks couple both to muons and to taus, it becomes possible to reduce the discrepancy between predictions and observations for all flavour anomalies, as well as to avoid the violation of any other constraints such as those arising from direct and indirect leptoquark searches at the LHC, from precision measurements at the $Z$-pole at LEP or from various other flavour observables~\cite{Becirevic:2018afm}\footnote{An extension of this model has been recently proposed as an explanation for the theory-experiment discrepancy related to the anomalous magnetic moment of the muon and the neutrino masses~\cite{Babu:2020hun}.}. Following the updated fit of ref.~\cite{Becirevic:2020pc}, we consider scenarios in which the leptoquark masses are $m_{R_2} = 1.3$\,TeV and $m_{S_3} = 2$\,TeV. All Yukawa couplings except those appearing in table~\ref{tab:benchmarksR2plusS3} are set to zero.

\paragraph{The singlet-triplet leptoquark model: $S_1$ and $S_3$}\label{sec:benchms1s3}
\renewcommand{\arraystretch}{1.5}\setlength\tabcolsep{5pt}
\begin{table}
\centering
 \begin{tabular}{l | cccc | cc | cccc}
    & $y_{1,22}^{LL}$ & $y_{1,23}^{LL}$ & $y_{1,32}^{LL}$ & $y_{1,33}^{LL}$ & $y_{1,23}^{RR}$ & $y_{1,32}^{RR}$ & $y_{3,22}^{LL}$ & $y_{3,23}^{LL}$ & $y_{3,32}^{LL}$ & $y_{3,33}^{LL}$
  \tabularnewline
  \hline
  $c_1$ & $-0.0082$ & $-1.46$ & $-0.016$ & $-0.064$ & $1.34$ & $-0.19$ & $-0.019$ & $0.58$ & $-0.059$ & $-0.11$
  \tabularnewline
  $c_2$ & $0.0078$ & $1.36$ & $-0.055$ & $0.052$ & $-1.47$ & $-0.053$ & $-0.017$ & $-1.23$ & $-0.070$ & $0.066$
  \tabularnewline
 \end{tabular}
 \caption{Two-leptoquark explanations to the flavour anomalies when the Standard Model is extended with $S_1$ and $S_3$ leptoquark species. We consider two scenarios $c_1$ and $c_2$ originating from the scan~\cite{Crivellin:2019dwb}. Our $c_1$ and $c_2$ setups correspond to the $p_1$ and $p_2$ benchmarks in the notation of ref.~\cite{Crivellin:2019dwb}. All Yukawa couplings not present in the table are set to zero, and the leptoquark masses read $m_{S_1} = m_{S_3} = \mlq = 1.2$\,TeV.}
 \label{tab:benchmarksS1plusS3}
\end{table}

As the last benchmark in our study, we consider the singlet-triplet model introduced in ref.~\cite{Crivellin:2017zlb,Crivellin:2019dwb}, and in which the Standard Model is extended by both an $S_1$ and an $S_3$ leptoquark. By considering leptoquark couplings to muons and taus, such a framework has been shown to provide an explanation for three of the most prominent flavour anomalies to date together. It not only yields an explanation for the $R_{K^{(*)}}$ and $R_{D^{(*)}}$ anomalies, but also allows for the reduction of the gap between the theoretical predictions and the observations of the anomalous magnetic moment of the muon. At the same time, this can be realised  without having to violate any bound that could emerge from LHC searches for leptoquark pair and single production.

The study~\cite{Crivellin:2019dwb} first highlighted 350 benchmark points satisfying all constraints. The authors next singled out four of these points that they finally labeled as $p_1$, $p_2$, $p_3$ and $p_4$\footnote{In a recent update of their analysis, the authors of~\cite{Crivellin:2019dwb} introduced new benchmark points $p_5$ to $p_8$. As in terms of leptoquark pair-production cross sections they do not lead to predictions that are significantly different from those for scenarios $p_1$ to $p_4$, we do not include them in our study.}. In the present study, we restrict our analysis to the points $p_1$ and $p_2$, that we re-label $c_1$ and $c_2$. In the results presented in this paper, we have ignored the points $p_3$ and $p_4$, as in terms of leptoquark pair-production cross sections the $p_1$ and $p_4$ points lead to very similar results, as do the points $p_2$ and $p_3$. We thus set both leptoquark masses to $m_{S_1} = m_{S_3} = \mlq = 1.2$\,TeV, and turn on the elements of the various Yukawa coupling matrices shown in table~\ref{tab:benchmarksS1plusS3} (all other entries to these matrices being once again taken vanishing).

\subsection{Precision calculations for scalar leptoquark pair production}\label{sec:lqprod}
\subsubsection{Scalar leptoquark pair production at fixed order in perturbation theory}

We discuss the production of a pair of scalar leptoquarks at hadron colliders, and at the LHC in particular,
\begin{equation}
	pp \to \lqp + X,
\end{equation}
where LQ generically stands for any species of considered leptoquarks. 

The fixed-order cross section $\sigma^{\text{NLO w/} t\text{-channel}}$, studied in this work, consists of the Born part $\sigma^{(0)}$ and  the QCD corrections  to the Born cross section $\sigma^{(1)}$. Taking into account the fact  that the Born amplitude contains contributions of a pure QCD nature (thus proportional to the strong coupling $\alphas$) and contributions that are $t$-channel-like (thus proportional to the leptoquark Yukawa coupling squared $y^2$), a power-counting of the couplings reveals the following contributions to  $\sigma^{\text{NLO w/} t\text{-channel}}$:
\begin{center}
	\begin{tabular}{rcccc}
                  & $ \sigma^{\text{NLO w/} t\text{-channel}}$ =&      $ \sigma^{(0)}$           & + &  $\sigma^{(1)}$ \\
		(\emph{1}): & & $\mathcal{O}(\alphas^2)$ & & $\mathcal{O}(\alphas^3)$;\\
		(\emph{2}): & & $\mathcal{O}(y^4)$ & & $\mathcal{O}(y^4 \alphas)$;\\
		(\emph{3}): & & $\mathcal{O}(y^2\alphas)$ & & $\mathcal{O}(y^2\alphas^2)$.
	\end{tabular}
\end{center}

A selection of representative Feynman diagrams is shown in figure~\ref{th:feynmandiagrams}. The contributions of class (\emph{1}) include the pure QCD contributions, corresponding to the purely QCD-mediated diagrams (a)--(c) shown in the top row of figure~\ref{th:feynmandiagrams}, whereas those of class (\emph{2}) refer to the contributions originating from lepton $t$-channel exchange as well as the QCD corrections to these diagrams, {\it c.f.}\ the diagrams (d)--(f) of the bottom row of figure~\ref{th:feynmandiagrams}. The terms of class (\emph{3}) consist of the additional mixed-order components induced by the interference of the QCD and $t$-channel diagrams of ${\cal O}(y^2\alphas)$ and ${\cal O}(y^2\alphas^2)$ at LO and NLO respectively. 

Previous precision calculation studies for scalar leptoquark pair production rely either on the NLO-QCD computations~\cite{Kramer:1997hh,Kramer:2004df}, {\it i.e.}\ the contributions of class (\emph{1}), or on their matching with parton showers~\cite{Mandal:2015lca,Dorsner:2018ynv}. In this paper, following our earlier work~\cite{Borschensky:2020hot}, we additionally include the $t$-channel lepton exchange contributions and their interference with QCD the diagrams, {\it i.e.}\ the contributions of class (\emph{2}) and class (\emph{3}). The sum of all three classes of contributions (\emph{1}+\emph{2}+\emph{3}) corresponds to our complete NLO-accurate prediction and is collectively coined ``NLO w/ $t$-channel'' in the following. This contrasts with the pure QCD case (\emph{1}) that we refer to as the ``NLO-QCD'' predictions.

\subsubsection{Resummation of soft-gluon corrections for scalar leptoquark pair production}
\label{sec:resummation}
Apart from considering the $t$-channel contributions and the NLO-QCD corrections to the complete set of LO diagrams, theory predictions can be improved by adding corrections due to soft-gluon emission. These  purely QCD contributions manifest themselves in the perturbative expansion of a cross section as logarithmic terms of the form
\begin{align}
	\alphas^n\ln^k\beta^2 \qquad \text{for}\ k\leq 2n,
\end{align}
where $\beta = \sqrt{1-4\mlq^2/s}$ is the relative velocity between the two produced leptoquarks of mass $\mlq$. The kinematical region close to the production threshold of the leptoquark pair, in which the partonic centre-of-mass energy $s\sim (2 \mlq)^2$, corresponds to the limit where only soft gluons can be emitted.  The terms logarithmic  in $\beta$ can then become large and need to be systematically taken into account, {\it i.e.}~resummed to all orders in perturbation theory.

In the following we resum those soft-gluon corrections to NNLL accuracy, and then match the NNLL results with NLO w/ $t$-channel predictions. NNLL resummation is performed in an analogous manner to calculations for squark and gluino production \cite{Kulesza:2008jb, Kulesza:2009kq, Beenakker:2009ha, Beenakker:2010nq, Beenakker:2011sf, Beenakker:2014sma}, and in particular for stop production~\cite{Beenakker:2016gmf}. However, for completeness, the threshold resummation formalism that we employ here is briefly reviewed in the rest of this section.

The inclusive hadronic cross section for the production of a leptoquark pair is written as
\be\bsp
	\sigma_{h_1h_2\to \lqp}&(\rho,\mlq,\mu_R,\mu_F) = \sum_{i,j}\int {\rm d}x_1\ {\rm d}x_2\ {\rm d}\hat{\rho}\ \delta\left(\hat{\rho}-\frac{\rho}{x_1x_2}\right)\\
	& \times f_{i/h_1}(x_1,\mu_F)\,f_{j/h_2}(x_2,\mu_F)\,\sigma_{ij\to \lqp }(\hat{\rho},\mlq,\mu_R,\mu_F),
\esp\ee
where $\rho$ is the hadronic threshold variable that measures the distance from the hadronic threshold. It is defined as $\rho \equiv 4\mlq^2/S$ with $S$ being the hadronic centre-of-mass energy of the collider. Moreover, $\sigma_{ij\to \lqp }(\hat{\rho},\mlq,\mu_R, \mu_F)$ is the partonic cross section. $f_{i/h_1}(x_1,\mu_F)$ and $f_{j/h_2}(x_2,\mu_F)$ are the parton distribution functions with $i$ and $j$ indicating the initial-state parton flavours, and $x_1$ and $x_2$ the momentum fractions of the partons inside the hadrons $h_1$ and $h_2$. The renormalisation and factorisation scales are denoted as $\mu_R$ and $\mu_F$, respectively.

Threshold resummation is carried out in Mellin-moment space, with the Mellin transform of the cross section given by
\be\bsp
	\tilde{\sigma}_{h_1h_2\to\lqp}&(N,\mlq,\mu_R,\mu_F) = \int_0^1 {\rm d}\rho\ \rho^{N-1}\ \sigma_{h_1h_2\to\lqp}(\rho,\mlq,\mu_R,\mu_F)\\
	&= \sum_{i,j}\tilde{f}_{i/h_1}(N+1,\mu_F)\,\tilde{f}_{j/h_2}(N+1,\mu_F)\,\tilde{\sigma}_{ij\to\lqp}(N,\mlq,\mu_R,\mu_F).
\esp\ee
The logarithmically enhanced terms are now of the form $\alphas^n\ln^k N$ with $k\leq 2n$, and the threshold limit $\beta\to 0$ corresponds to $N\to\infty$. The all-order summation of such logarithmic terms follows from the near-threshold factorisation of the partonic cross section into functions that describe the different kinds of gluon emission: hard radiation, collinear radiation, and wide-angle soft radiation~\cite{Sterman:1986aj,Catani:1989ne,Bonciani:1998vc,Contopanagos:1996nh,Kidonakis:1998bk,Kidonakis:1998nf}. In terms of these functions the partonic cross section can be written near threshold as
\begin{align}
 \tilde \sigma^{\mathrm{res, NNLL}}_{ij\rightarrow \lqp,I}(N) = \tilde\sigma^{(0)}_{ij\rightarrow \lqp,I}(N)\,\tilde C_{ij\rightarrow \lqp,I}(N)\,\Delta^{S}_I(N+1)\,\Delta_i(N+1)\,\Delta_j(N+1), 
\label{eq:partsigN}
\end{align}
where, in order to keep the notation compact, we only show the dependence on a single argument, the Mellin variable $N$. The index $I$  denotes the colour representation of the final state, {\it i.e.}\ either a singlet ($I={\mathbf 1}$) or an octet ($I={\mathbf 8}$). The cross section $\tilde{\sigma}^{(0)}_{ij\to \lqp,I}$ is the Born cross section in Mellin-moment space projected onto the colour state $I$. The explicit expression for the incoming jet radiative factors $\Delta_i$ and the soft emission factor $\Delta^{S}_I$ can be found, {\it e.g.},\ in~\cite{Beenakker:2016gmf}. Their product is given, at the NNLL accuracy, by
\begin{align}
\label{eq:deltas}
 \Delta^{S}_I (N)\,\Delta_i (N)\, \Delta_j (N)=\exp{\Big[L\,g_1{(\alphas L)}+g_2{(\alphas L)}+\alphas g_3{(\alphas L)}\Big]}.
\end{align}
This exponential function resums the logarithms $L = \ln N$ originating from soft-collinear gluon radiation. The function $g_1$ provides the leading logarithmic (LL) approximation, while the inclusion of $g_2$ and $g_3$ accounts for the next-to-leading-logarithmic (NLL) and the NNLL contributions respectively. Expressions for the functions $g_1$ and $g_2$ can be found, for example, in ref.~\cite{Kulesza:2009kq}, while $g_3$ is available in, {\it e.g.}, ref.~\cite{Beenakker:2011sf}. The hard matching coefficient $\tilde C_{ij\to \lqp,I}$ appearing in eq.~(\ref{eq:partsigN}) contains higher-order terms that do not vanish in the threshold limit, and that are different from the logarithms included in eq.~(\ref{eq:deltas}). As shown in \cite{Beneke:2010da}, close to threshold these terms factorise into off-shell hard contributions and Coulomb contributions originating from exchanges of gluons between two slowly-moving coloured particles in the final state. Since up to NNLL precision we only need to know the hard matching coefficient up to ${\cal O}(\alphas)$, we use
\begin{align}
 C_{ij\rightarrow \lqp ,I} = 1 + \frac{\alphas}{\pi} {\cal C}^{(1)}_{ij\rightarrow \lqp,I} + \frac{\alphas}{\pi} {\cal C}^{\text{Coul},(1)}_{I}
\end{align}
as an input for
\begin{align}
 \tilde\sigma^{(0)}_{ij\rightarrow \lqp ,I}(N)\,\tilde C_{ij\rightarrow \lqp,I}(N) = \int \text{d} \hat\rho \, \hat \rho^{N-1}\, \sigma^{(0)}_{ij\rightarrow \lqp,I}(\hat \rho)\,C_{ij\rightarrow \lqp,I}(\hat \rho),
\label{eq:hardmatching_mellin}
\end{align}
where only pure QCD contributions are considered. The one-loop Coulomb contributions are well known and read in our case
$$
{\cal C}^{\text{Coul},(1)}_{I} = \frac{\alphas \pi}{2 \beta} \kappa_I    \qquad\text{with}\qquad \kappa_{\mathbf 1}= \frac{4}{3}\ \ \text{and}\ \ \kappa_{\mathbf 8}=-\frac{1}{6},
$$
whereas the remaining contributions ${\cal C}^{(1)}_{ij\rightarrow \lqp,I}$ are obtained from the analogous result for stop pair production~\cite{Beenakker:2016gmf}\footnote{In the decoupling limit of large squark and gluino masses, the only difference between the hard matching coefficients for leptoquark pair and stop pair production originates from the four-stop vertex. This difference can be however safely neglected as the four-stop vertex contributions consist of a permille-level correction for the final state masses considered here.}.

In order to combine the resummed QCD predictions together with the fixed-order calculations, a matching procedure is needed to prevent the double-counting of contributions present both in the resummed and fixed-order results. This matched cross section reads
\be\bsp
	&\sigma^{\text{NLO w/} t\text{-channel + NNLL}}_{h_1h_2\to \lqp}(\rho,\mlq,\mathbf{y},\mu_R,\mu_F) = \sigma^{\text{NLO w/} t\text{-channel}}_{h_1h_2\to\lqp}(\rho,\mlq,\mathbf{y},\mu_R,\mu_F)\\
	&\qquad\qquad+\sum_{i,j}\int_{\cal{C}}\frac{dN}{2\pi i}\rho^{-N}\tilde{f}_{i/h_1}(N+1,\mu_F)\,\tilde{f}_{j/h_2}(N+1,\mu_F)\\
	&\qquad\qquad\times\left[\tilde{\sigma}^{\mathrm{res,NNLL}}_{ij\to\lqp}(N,\mlq,\mu_R,\mu_F)-\tilde{\sigma}^{\mathrm{res,NNLL}}_{ij\to\lqp}(N,\mlq,\mu_R,\mu_F)\big|_{\mathrm{NLO}}\right].
\esp\label{invmellin}\ee
In this expression, $\tilde{\sigma}^{\mathrm{res,NNLL}}_{ij\to\lqp}(N,\mlq,\mu_R,\mu_F)\big|_{\mathrm{NLO}}$ stands for the expansion of the resummed cross section up to NLO in Mellin-moment space. Subtracting this expansion from the resummed cross section prevents the double counting of terms already taken into account in the fixed-order part $\sigma^{\text{NLO w/} t\text{-channel}}_{h_1h_2\to\lqp}$. Finally, an inverse Mellin transform allows for the derivation of cross section predictions in physical momentum space. This transform is carried out by integrating along a contour $\cal{C}$ according to the “minimal prescription”~\cite{Catani:1996yz}.

\section{Numerical implementation and set-up}\label{sec:numerics}
In the present section, we discuss the implementations of the considered leptoquark model that are at the heart of our precision computations. We make use of two general frameworks for higher-order calculations matched with parton showers, namely the \mgamc~\cite{Alwall:2014hca} and \pwb{}~\cite{Nason:2004rx,Frixione:2007vw,Alioli:2010xd} frameworks. In both cases, external quark, leptoquark, and gluon fields are renormalised on-shell, while for the couplings and the masses we use the $\MSbar$ scheme. For the running of the strong coupling constant $\alphas$, the heavy states such as the leptoquark fields as well as the top quark have been decoupled. This corresponds to a subtraction of the logarithms which originate from the leptoquark and top quark contributions to the vacuum polarisation of the gluon, that are then absorbed in the renormalisation constant of the strong coupling~\cite{Collins:1978wz,Bardeen:1978yd,Marciano:1983pj}. We thus recover a five-flavour-scheme running as in the Standard Model. Our calculations are performed with the CKM and PMNS matrices set to the identity.

Relying on two independent platforms enables us to ensure the correctness of our predictions. Moreover, we have checked the cancellation of both the ultraviolet and infrared divergences explicitly. Whilst both \mgamc\ and the \pwb\ would allow us to match NLO fixed-order calculations with parton showers, we restrain ourselves from doing so. Including the $t$-channel diagrams at the Born level leads to mixed-order contributions at NLO. Matching them with parton showers requires a general renormalisation procedure (including counterterms at ${\cal O}(\alphas)$ as well as at ${\cal O}({\bf y})$), a computation of all associated $R_2$ rational terms relevant for the numerical evaluation of loop amplitudes~\cite{Ossola:2006us,Ossola:2007ax}, and the derivation of all necessary Frixione-Kunszt-Signer (FKS) subtraction terms relevant to treat the inherent infrared divergences~\cite{Frixione:1995ms}. Such tasks go well beyond the scope of this work that solely targets improving theoretical predictions for leptoquark pair-production total rates.

\subsection{\mgamc\ implementation}
In order to derive NLO fixed-order results and extract the different components relevant for resummation at the NNLL accuracy, we first implement the simplified model introduced in section~\ref{sec:model} into \fr~\cite{Alloul:2013bka}. The bare Lagrangian~\eqref{eq:lag} is then renormalised in the on-shell scheme and at ${\cal O}(\alphas)$ with the help of NLOCT~\cite{Degrande:2014vpa} and \fa~\cite{Hahn:2000kx}. Next, we generate a corresponding UFO library~\cite{Christensen:2009jx,Degrande:2011ua} that enables us to perform cross section calculations through the \mgamc\ platform~\cite{Alwall:2014hca} for processes involving any of the leptoquark species considered. This however requires some modifications in \mgamc, since in the presence of leptoquarks loop diagrams containing leptons should not be systematically vetoed (as done automatically for NLO-QCD calculations in \mgamc). Their inclusion is achieved by making use of hard-coded loop filters in the files \texttt{base\_objects.py} and \texttt{loop\_diagram\_generation.py}. Further details on the required modifications can be found in appendix~\ref{app:mg}.

\subsection{\pwb{} implementation}
In order to cross-check the results that have been obtained with the \mgamc{} code, a second calculation of the NLO-QCD corrections has been implemented independently within the \pwb{} framework~\cite{Nason:2004rx,Frixione:2007vw,Alioli:2010xd}. The main ingredients that are required for the implementation of the calculations performed in this work consist of the Born, virtual, and real-emission matrix elements.

The Born and real-emission parts are automatically generated by a tool based on \mgf{}~\cite{Murayama:1992gi,Stelzer:1994ta,Alwall:2007st}, for which we have implemented a model based on the simplified framework of section~\ref{sec:model}. Our implementation includes, in particular, the lepton-quark-leptoquark interactions which lead to the additional $t$-channel contributions discussed in this work. The automated tool also provides spin- and colour-correlated Born amplitudes which are required to construct additional terms necessary for the subtraction of infrared divergences in the FKS scheme \cite{Frixione:1995ms}.

In the \pwb{} code, the routines \texttt{is\_charged}, \texttt{is\_coloured}, and \texttt{btildevirt} have been modified to correctly include the electric and colour charges of the leptoquark states and assign the proper corresponding colour factors, as detailed in appendix~\ref{app:pwb}. For the virtual amplitudes, we have manually implemented the interaction vertices of the discussed model in a \fa{}~\cite{Hahn:2000kx} model file, which is then used in conjunction with \fc{}~\cite{Hahn:1998yk} to compute the necessary one-loop matrix elements, as well as the relevant renormalisation constants. For the numerical evaluation of the one-loop integrals, we interface our code with the library \collier{}~\cite{Denner:2016kdg,Denner:2002ii,Denner:2005nn,Denner:2010tr}. 

\subsection{Resummation at the next-to-next-to-leading logarithmic accuracy}

The threshold corrections resummed in this work up to the NNLL accuracy account for pure QCD effects from soft-gluon emission and as such do not depend on the details of a particular leptoquark model. As a consequence,  the resummation approach applied here bears a resemblance to calculations performed for squark and gluino pair production \cite{Kulesza:2008jb, Kulesza:2009kq, Beenakker:2009ha, Beenakker:2010nq, Beenakker:2011sf, Beenakker:2014sma}, and in particular for top squark pair production \cite{Beenakker:2016gmf} (see the discussion in section~\ref{sec:resummation}). The already exiting numerical package dedicated to resummation calculations for this class of processes, {\textsc{NNLL-fast}} \cite{Beenakker:2016lwe}, can therefore be used. In order to extend the {\textsc{NNLL-fast}} platform to include scalar leptoquark pair production, two independent in-house resummation codes have been developed.

\section{Precision predictions for scalar leptoquark pair production}\label{sec:results}
In this section, we extend the discussion in our previous letter~\cite{Borschensky:2020hot} in which we have briefly investigated several aspects of NLO w/ $t$-channel+NNLL predictions for scalar leptoquark pair production at hadron colliders. We comprehensively study the set of associated effects for specific leptoquark scenarios (see section~\ref{sec:benchmarks}) that have been chosen in the light of the recent flavour anomalies. All calculations correspond to a centre-of-mass energy $\sqrt{S} = 13$ TeV, and hadronic cross sections are evaluated by convoluting partonic cross sections with three different sets of parton distribution functions (PDFs), namely CT18~\cite{Hou:2019efy}, NNPDF3.1~\cite{NNPDF:2017mvq}, and MSHT20~\cite{Bailey:2020ooq}. Unless stated otherwise, we employ NLO PDF sets for predictions including pure NLO-QCD and NLO w/ $t$-channel corrections, while NNLO sets are used for predictions at the NLO+NNLL level\footnote{As NNLL resummation is expected to capture the bulk of the NNLO corrections, we consider that employing NNLO PDF sets is more suitable than employing NLO PDF sets for predictions including NNLL corrections.}. As central renormalisation and factorisation scales, we use $\mu_R = \mu_F = \mlq$. The scale uncertainties are then estimated through the seven-point method, in which the renormalisation and factorisation scales are varied independently by a factor of 2 up and down relative to their central value, with the configurations where the two scales are the furthest apart being left out.

In section~\ref{sec:results:individualimpact}, we consider simple scenarios in which the Standard Model is extended by a single leptoquark species (as described in section~\ref{sec:benchmsimple}), whereas more realistic models (introduced in sections~\ref{sec:benchmr2}, \ref{sec:benchmr2s3} and \ref{sec:benchms1s3}) are studied in section~\ref{sec:results:benchmarks}. In section~\ref{sec:exp_impact}, we investigate the potential impact of our calculations on limits derived by the LHC collaborations.

\subsection{Relative importance of the $t$-channel contributions, soft-gluon resummation, and the parton densities}\label{sec:results:individualimpact}

\begin{figure}
\centering
\includegraphics[width=0.55\textwidth]{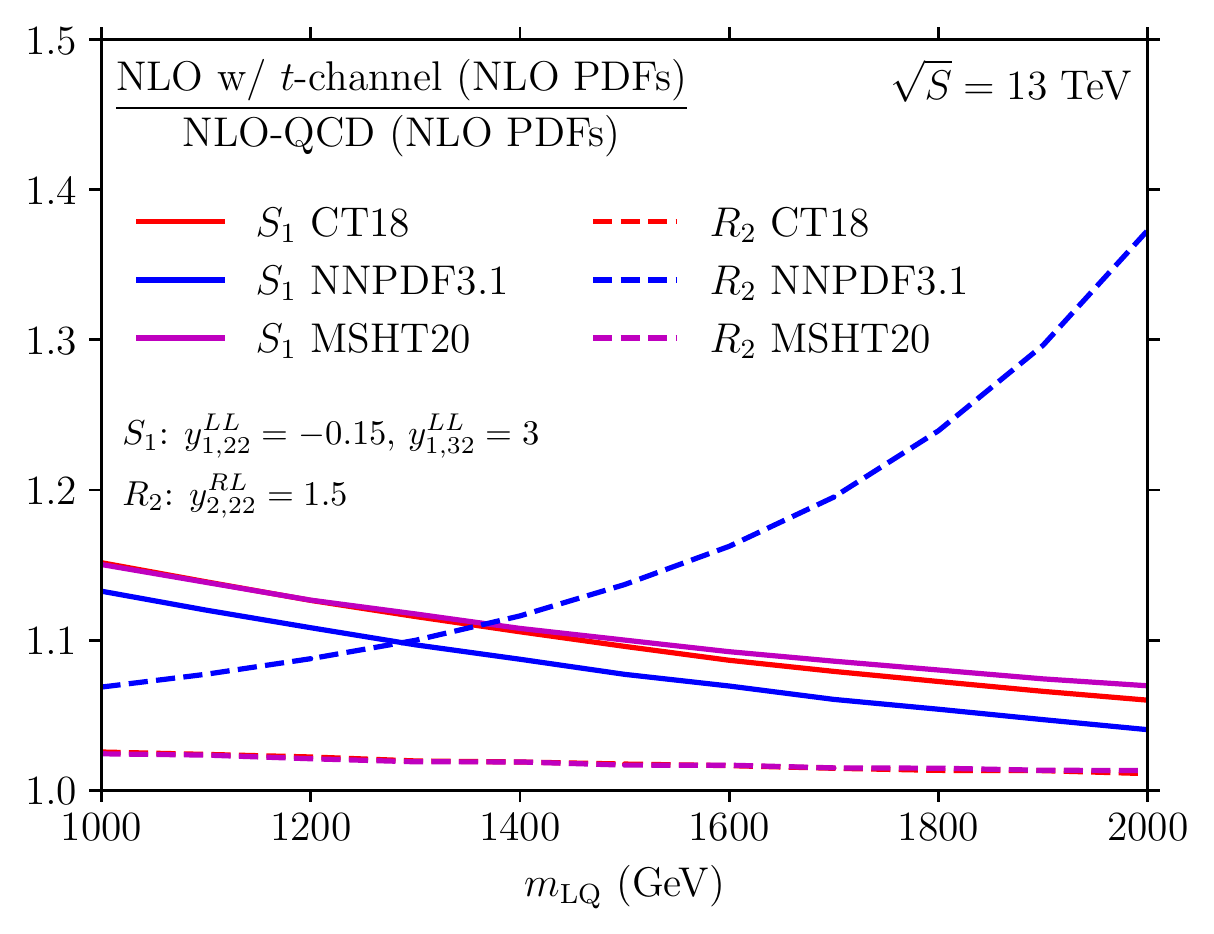}
\caption{Impact of the $t$-channel contributions on the total cross section associated with $S_1^{(-1/3)} S_1^{(+1/3)}$ (solid) and $R_2^{(+5/3)} R_2^{(-5/3)}$ (dashed) production. The results are shown as ratios of NLO w/ $t$-channel over pure NLO-QCD predictions, and we present them for the CT18 (red), NNPDF3.1 (blue), and MSHT20 (magenta) PDF sets.}
\label{fig:lq:tchannel-contributions}
\end{figure}

In order to examine the relative importance of the corrections under consideration in this work, we begin with the analysis of their individual impact on the predictions. For this purpose we omit the calculation of PDF uncertainties. This will be performed in the next subsections. We consider the simple scenarios of section~\ref{sec:benchmsimple} in which the Standard Model is extended by a single leptoquark species, and we discuss the pair-production processes $pp\to S_1^{(-1/3)} S_1^{(+1/3)}$ and $pp\to R_2^{(+5/3)} R_2^{(-5/3)}$.

In figure~\ref{fig:lq:tchannel-contributions}, we assess how the $t$-channel contributions affect NLO predictions by studying the ratio $K_{\rm wt}$ of NLO cross sections
\be\label{eq:Kwt}
  K_{\rm wt} \equiv \frac{\sigma^{\text{NLO w/} t\text{-channel}} (\text{NLO PDFs})}{\sigma^{\text{NLO-QCD}}(\text{NLO PDFs})},
\ee
for different choices of NLO parton density sets and different values of the leptoquark mass $\mlq$. The additional $t$-channel contributions are always positive for the considered mass range (as $K_{\rm wt} > 1$). However, the behaviour of these corrections depends strongly on the leptoquark nature (and thus on the structure of its Yukawa interactions), and on the PDF set that is used for the numerical computations. For $S_1$ pair production, the impact of the $t$-channel contributions is similar for all three PDF sets and get smaller with increasing $\mlq$ values (solid curves). It hence yields an increase of the cross section with respect to NLO-QCD predictions that ranges from 15\% to 4\% for leptoquark masses $\mlq$ varying from 1 to 2 TeV respectively. In contrast, we obtain a very mild and constant cross section increase of 1\% or 2\% for $R_2$ pair production when we use the CT18 and MSHT20 parton densities (dashed red and magenta curves), whereas employing NNPDF3.1 densities leads to a growth in cross section ranging from 7\% for small leptoquark mass values to about 37\% for $\mlq=2$~TeV (dashed blue curve). Such a drastically different behaviour can be traced back to the adopted texture for the ${\bf y_{\sss 2}^{\sss RL}}$ matrix in the considered $R_2$ simple benchmark, leading to important contributions from $c\bar c$-initiated subprocesses. The different behaviour of the predictions relying on the NNPDF3.1 set originates then from its different parametrisation of the charm PDF, which is fitted alongside the quark and gluon densities and not generated perturbatively as in the case of the CT18 and MSHT20 sets.

\begin{figure}
\centering
\includegraphics[width=0.55\textwidth]{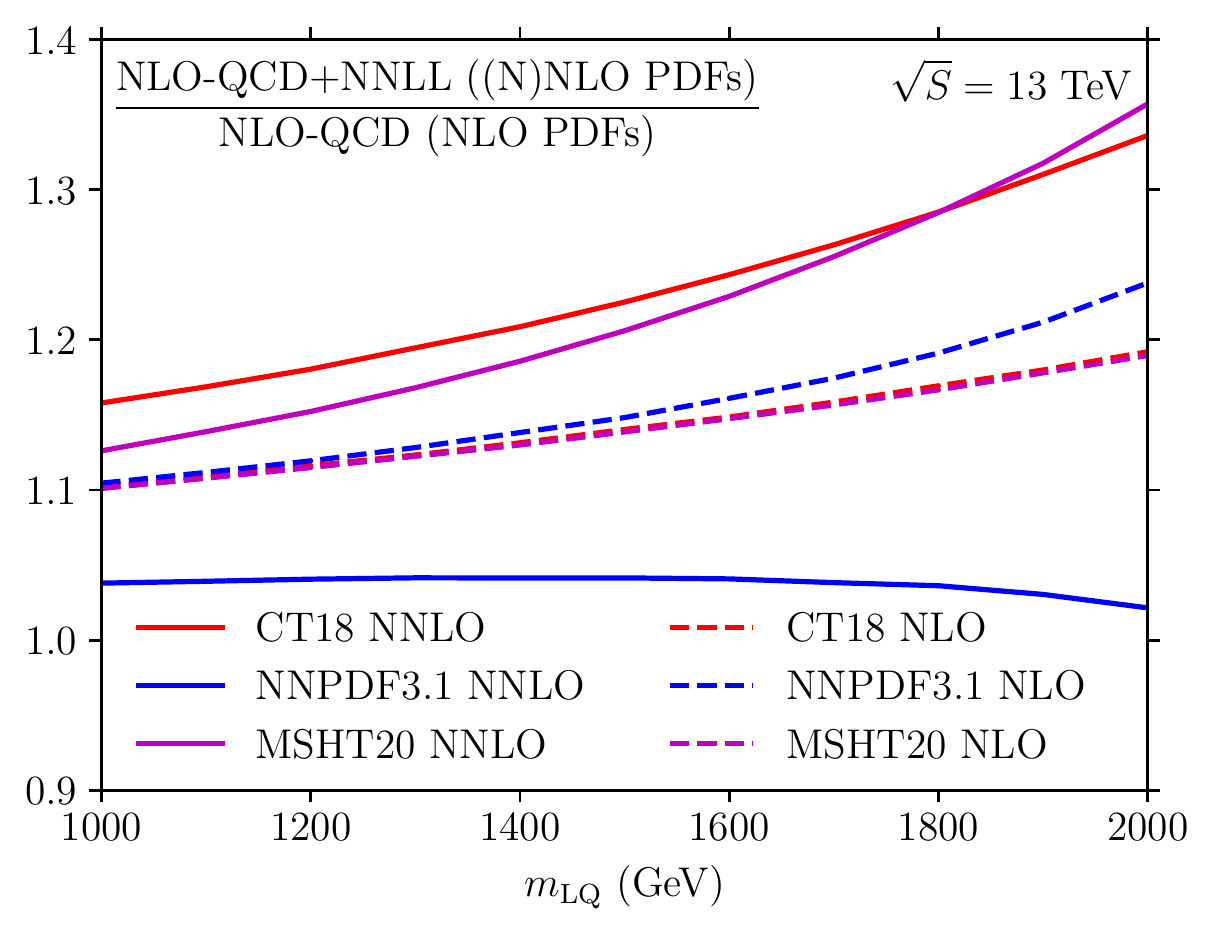}
\caption{Impact of soft-gluon resummation corrections on leptoquark pair-production cross sections. The predictions are shown as ratios of NLO-QCD+NNLL over NLO-QCD cross sections, for the CT18 (red), NNPDF3.1 (blue), and MSHT20 (magenta) PDF sets. We turn off the $t$-channel contributions and additionally assess the effect of using NLO (dashed) instead of NNLO (solid) parton densities.}
\label{fig:lq:qcdresum-contributions}
\end{figure}

In figure~\ref{fig:lq:qcdresum-contributions}, we study the impact of soft-gluon resummation corrections that we match with the pure NLO-QCD results in this plot. To this end, we calculate the ratio $K_{\rm QCD}$ of the NLO+NNLL cross section to the NLO one,
\be
  K_{\rm QCD} \equiv \frac{\sigma^{\text{NLO-QCD+NNLL}}(\text{(N)NLO PDFs})}{\sigma^{\text{NLO-QCD}}(\text{NLO PDFs})}.
\ee
We investigate the dependence of $K_{\rm QCD}$ on the leptoquark mass, and on different choices of NLO and NNLO parton densities. As $t$-channel contributions are not relevant in this case, only the leptoquark mass plays a role and the results are independent of the leptoquark Yukawa couplings and their nature. While resummation corrections are always positive, the chosen PDF fit has a large impact on their relative size. We compute predictions with NLO (dashed curves) and NNLO (solid curves) PDF sets for the NLO-QCD+NNLL results, whereas NLO sets are always used for the NLO-QCD rates appearing in the denominator of $K_{\rm QCD}$. A good agreement is found when NLO PDFs are used, all predictions yielding a cross section increase ranging from 10 to 24\% for leptoquark masses varying in the [1, 2]~TeV range. The situation is however quite different when NNLO PDFs are considered. In this case, NNPDF3.1 predictions (blue) are significantly smaller than the CT18 and MSHT20 ones (red and magenta). In the former case we find a mild cross section increase of 2\% to 4\%, whereas in the latter case the cross section strongly grows with the mass, the increase ranging from 13\% to 34\%. This different behaviour is again related to the different treatment of the charm density in NNPDF3.1, this time at NNLO, which perturbatively interplays with all other parton densities. The relevant gluon, up and antiup NNLO PDFs are in particular quite affected by this for Bjorken $x$ in the 0.1--0.5 range.

\begin{figure}
\centering
\includegraphics[width=0.55\textwidth]{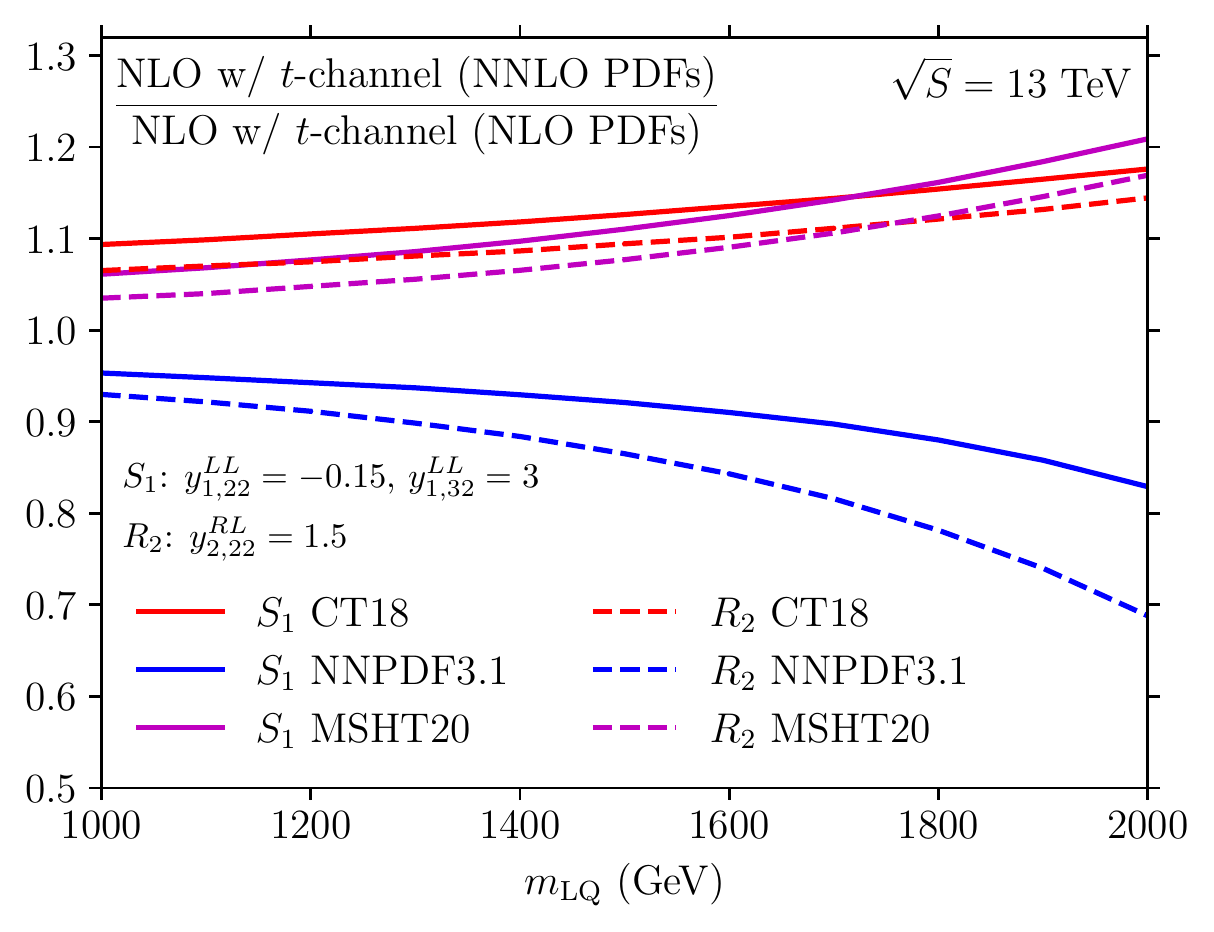}
\caption{Impact of the choice between an NLO and an NNLO PDF set on $S_1^{(-1/3)} S_1^{(+1/3)}$ (solid) and $R_2^{(+5/3)} R_2^{(-5/3)}$ (dashed) production. The predictions are shown as ratios of NLO w/ $t$-channel cross sections (without any resummation corrections) computed with NNLO parton densities over the rates correspondingly obtained with NLO PDF sets. We consider the CT18 (red), NNPDF3.1 (blue), and MSHT20 (magenta) densities.}
\label{fig:lq:pdf-contributions}
\end{figure}

This different behaviour resulting from the differences between NLO and NNLO parton densities is further investigated in figure~\ref{fig:lq:pdf-contributions}, now in the context of the $t$-channel contributions. We display there the ratio $K_{\rm PDF}$ between NLO w/ $t$-channel cross sections computed with different parton densities. This ratio is defined as predictions relying on NNLO parton densities to those relying on NLO parton densities,
\be
  K_{\rm PDF} \equiv \frac{\sigma^{\text{NLO w/} t\text{-channel}} (\text{NNLO PDFs})}{\sigma^{\text{NLO w/} t\text{-channel}}(\text{NLO PDFs})}.
\ee
We study the dependence of $K_{\rm PDF}$ on the leptoquark mass for the two processes $pp\to S_1^{(-1/3)} S_1^{(+1/3)}$ and $pp\to R_2^{(+5/3)} R_2^{(-5/3)}$, and we present numerical results for all different parton density sets under consideration. Once again, NNPDF3.1 predictions (blue curves) are found to be quite different from all the other cases (red and magenta curves for CT18 and MSHT20 respectively). NNPDF3.1 results are found to be below 1, {\it i.e.}\ the use of NNLO PDFs leads to a reduction of the cross section. This decrease ranges from about $-5\%$ for light leptoquarks to up to $-31\%$ for heavier scenarios. On the contrary, making use of NNLO PDFs for cross section evaluations with either the CT18 or the MSHT20 set yields an increase of the rate of approximately 7\% to 20\%, the exact value depending on the process and on the leptoquark mass (the larger the mass, the more pronounced the increase).

\begin{figure}
\centering
\includegraphics[width=0.55\textwidth]{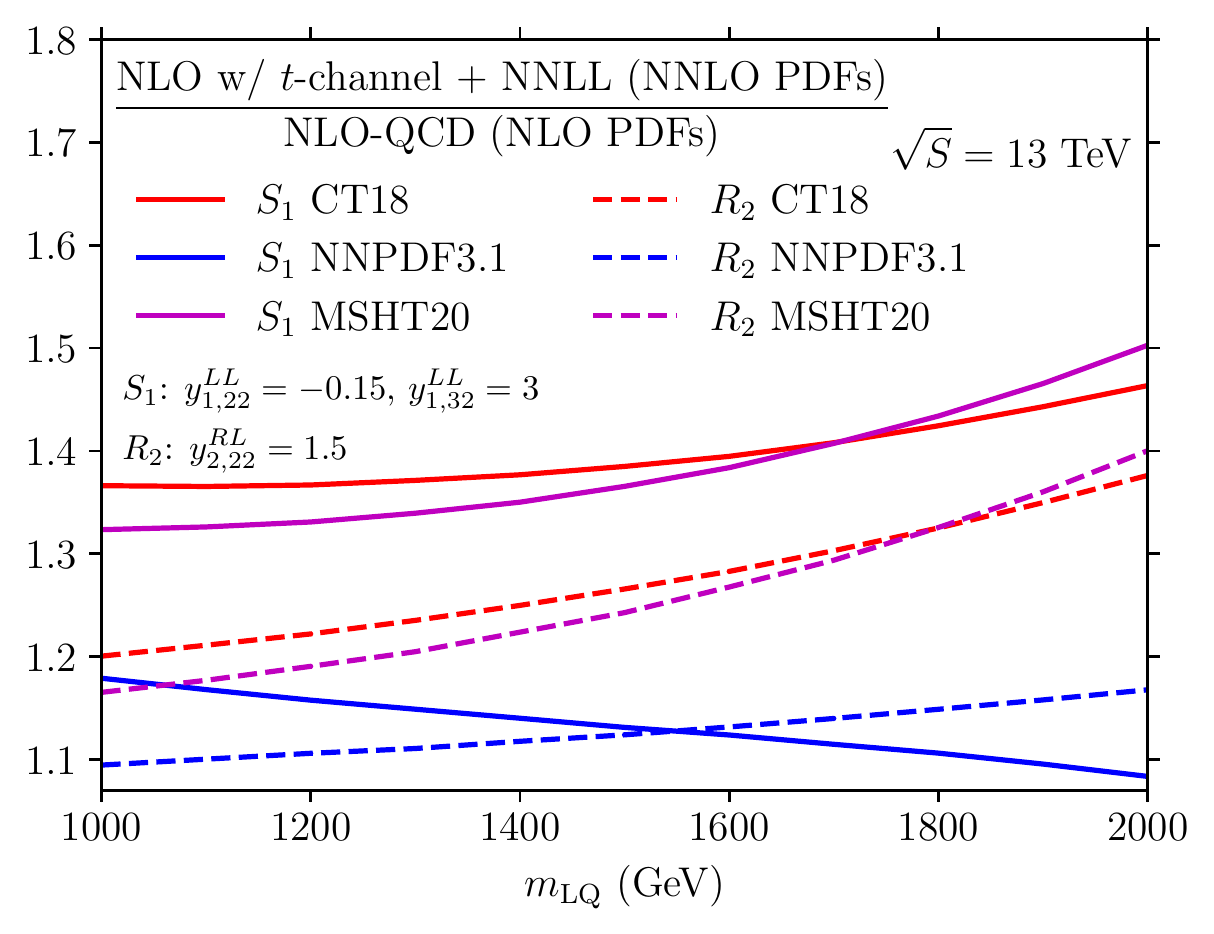}
\caption{Combined impact of all contributions studied in figures \ref{fig:lq:tchannel-contributions}--\ref{fig:lq:pdf-contributions} on $S_1^{(-1/3)} S_1^{(+1/3)}$ (solid) and $R_2^{(+5/3)} R_2^{(-5/3)}$ (dashed) production. The predictions are shown as ratios of the NLO w/ $t$-channel + NNLL cross sections computed with NNLO PDFs over the NLO-QCD cross sections evaluated with NLO PDF sets. We present results for the CT18 (red), NNPDF3.1 (blue), and MSHT20 (magenta) densities.}
\label{fig:lq:all-contributions}
\end{figure}

Finally, in figure~\ref{fig:lq:all-contributions} we combine all the effects studied in this work into a single ratio $K_{\rm NNLL}$ defined by
\be\label{eq:Knnll}
  K_{\rm NNLL} \equiv \frac{\sigma^{\text{NLO w/} t\text{-channel + NNLL}}(\text{NNLO PDFs})}{\sigma^{\text{NLO-QCD}}(\text{NLO PDFs})}.
\ee
In this figure, we present the dependence of $K_{\rm NNLL}$ on the leptoquark mass $\mlq$ and the chosen parton densities. Those ratios correspond to our best and most precise predictions, namely the NLO w/ $t$-channel + NNLL ones computed with NNLO PDF sets compared to the NLO-QCD predictions computed with NLO PDF sets. While the corrections with respect to the NLO-QCD rates are always positive ({\it i.e.}\ $K_{\rm NNLL}$>1), their actual size depends on a strong interplay between the leptoquark mass, the chosen PDF set, and the considered final state. The largest effects can be seen for the $S_1$ final state (solid curves), and for the CT18 (red curves) and MSHT20 (magenta curves) PDFs. The behaviour that is obtained when using CT18 or MSHT20 densities is generally similar, with corrections that are found to lie between approximately 30\% and 50\% in the shown mass range. For $R_2$ production (dashed curves), the size of the corrections stretches between 15\% and 40\% for both CT18 and MSHT20 densities. In contrast, cross sections obtained with NNPDF3.1 (blue curves) exhibit a very different behaviour, as it was already the case for the individual corrections considered. The size of the cross section growth is smaller than for the other two sets of parton densities, with the increase approximately between 10\% and 20\%. For the $S_1$ final state, the corrections get smaller with increasing leptoquark mass. An opposite behaviour is found for the $R_2$ process, for which the treatment of the charm density in the PDF fits is much more relevant by virtue of the adopted texture for the leptoquark Yukawa couplings.

Due to these large differences, it is therefore very important to consider all types of effects together, {\it i.e.}\ $t$-channel contributions, resummation corrections, as well as different choices for the PDF sets. Moreover, we caution the reader that the values of the $K$-factors given in tables~\ref{tab:benchmarksR2xsections}--\ref{tab:benchmarksS1plusS3xsections} are very strongly dependent on the considered model and its parameters. They can therefore be directly used only in analyses assuming exactly the same scenarios as considered in this paper. If this is not the case for a given phenomenological or experimental study, we advocate to re-weight the LO rate to NLO w/ $t$-channel+NNLL, the last predictions being evaluated directly for the benchnark scenario considered. To this aim, dedicated numerical packages are provided on the webpage \url{https://www.uni-muenster.de/Physik.TP/research/kulesza/leptoquarks.html}.

\subsection{Predictions for benchmark scenarios relevant for the flavour anomalies}\label{sec:results:benchmarks}
In the previous subsection we have studied the individual contributions that we take into account in our precision predictions and we have investigated their effects through ratios of cross sections. We now move on with the presentation of NLO w/ $t$-channel and NLO w/ $t$-channel+NNLL pair-production cross sections. We consider the more realistic scenarios that we have discussed in sections~\ref{sec:benchmr2}, \ref{sec:benchmr2s3}, and \ref{sec:benchms1s3}. We present in addition the theoretical uncertainties associated with our predictions.

\renewcommand{\arraystretch}{1.5}\setlength\tabcolsep{5pt}
\begin{table}
 \centering
 \begin{tabular}{l | l | l | lll | ll}
  & LQ & PDF & QCD (fb) & NLOwt (fb) & NNLLwt (fb) & $K_\text{wt}$ & $K_\text{NNLL}$
  \tabularnewline
  \hline
  \multirow{3}{*}{$a_1$} & \multirow{3}{*}{$R_2$} & CT18 & $5.49^{+11.4\%}_{-13.5\%}{}^{+11.4\%}_{-11.4\%}$ & $6.91^{+7.7\%}_{-9.8\%}{}^{+15.5\%}_{-15.5\%}$ & $8.71^{+4.6\%}_{-3.1\%}{}^{+22.5\%}_{-22.5\%}$ & $1.26$ & $1.59$
  \tabularnewline
  & & NNPDF3.1 & $5.24^{+11.6\%}_{-13.6\%}{}^{+3.8\%}_{-3.8\%}$ & $8.92^{+4.2\%}_{-6.1\%}{}^{+28.0\%}_{-28.0\%}$ & $8.40^{+4.2\%}_{-2.5\%}{}^{+30.2\%}_{-30.2\%}$ & $1.70$ & $1.60$
  \tabularnewline
  & & MSHT20 & $5.55^{+11.4\%}_{-13.4\%}{}^{+4.7\%}_{-4.7\%}$ & $6.88^{+8.0\%}_{-10.1\%}{}^{+6.6\%}_{-6.6\%}$ & $8.26^{+4.8\%}_{-3.3\%}{}^{+7.9\%}_{-7.9\%}$ & $1.24$ & $1.49$
  \tabularnewline
  \hline
  \multirow{3}{*}{$a_2$} & \multirow{3}{*}{$R_2$} & CT18 & $5.49^{+11.4\%}_{-13.5\%}{}^{+11.4\%}_{-11.4\%}$ & $5.51^{+11.5\%}_{-13.2\%}{}^{+11.5\%}_{-11.5\%}$ & $6.40^{+6.7\%}_{-6.1\%}{}^{+15.3\%}_{-15.3\%}$ & $1.00$ & $1.17$
  \tabularnewline
  & & NNPDF3.1 & $5.24^{+11.6\%}_{-13.6\%}{}^{+3.8\%}_{-3.8\%}$ & $5.30^{+11.4\%}_{-13.3\%}{}^{+4.0\%}_{-4.0\%}$ & $5.48^{+7.3\%}_{-5.9\%}{}^{+3.9\%}_{-3.9\%}$ & $1.01$ & $1.05$
  \tabularnewline
  & & MSHT20 & $5.55^{+11.4\%}_{-13.4\%}{}^{+4.7\%}_{-4.7\%}$ & $5.58^{+11.2\%}_{-13.4\%}{}^{+4.8\%}_{-4.8\%}$ & $6.29^{+6.7\%}_{-6.1\%}{}^{+5.3\%}_{-5.3\%}$ & $1.01$ & $1.13$
 \end{tabular}
 \caption{Total cross sections relevant for the $a_1$ and $a_2$ benchmark points of section~\ref{sec:benchmr2}. We consider $R_2$ pair production and evaluate the rates at the NLO-QCD (QCD), NLO w/ $t$-channel (NLOwt) and NLO w/ $t$-channel+NNLL (NNLLwt) accuracy. The first (second) error included in our results consists of the scale (symmetric PDF) uncertainties. Moreover, the cross sections for the two $R_2$ mass eigenstates are identical and thus are provided as single predictions for both benchmark points.}
 \label{tab:benchmarksR2xsections}
\end{table}

The scenarios that we have described in section~\ref{sec:benchmr2} are related to a leptoquark model that solely includes a weak doublet of $R_2$ leptoquarks. We have introduced two specific benchmarks $a_1$ and $a_2$ in table~\ref{tab:benchmarksR2}, those setups providing an explanation for the flavour anomalies. We present the associated cross section predictions in table~\ref{tab:benchmarksR2xsections} for the CT18, NNPDF3.1, and MSHT20 parton densities. The cross sections related to the two mass eigenstates $R_2^{(+5/3)}$ and $R_2^{(+2/3)}$ are equal in these two scenarios, since the decisive parameter is the $y^{RL}_{2,23}$ Yukawa coupling, {\it i.e.}\ the coupling of the corresponding leptoquark eigenstate to a charm quark and either a tau lepton or a tau neutrino. Both the third-generation charged lepton and neutrino are effectively massless in our calculations, so that the total cross section predictions are identical (and thus only provided once in the table).

The obtained values for the ratio $K_{\rm wt}$ defined in eq.~\eqref{eq:Kwt} illustrate the different behaviour of the $t$-channel contributions. As can be seen from table~\ref{tab:benchmarksR2xsections}, for benchmark point $a_1$ the ratio $K_{\rm wt}$ lies between $1.24$ and $1.70$, the exact value depending on the chosen PDF set. In contrast, the effects of the $t$-channel contributions are negligible in the case of scenario $a_2$. This behaviour is dominantly driven by the size of the $y^{RL}_{2,23}$ Yukawa coupling. For scenario $a_1$, this coupling is large relative to all other non-zero elements of the Yukawa mixing matrices, so that the $c\bar c$-initiated subprocesses significantly contribute. In comparison, for scenario $a_2$ all non-zero Yukawa couplings are much smaller, which suppresses automatically the relevance of any $t$-channel lepton exchange diagram. The NNPDF3.1 $K_{\rm wt}$ value for benchmark point $a_1$ is found to significantly differ from the values obtained when any of the two other PDF sets is considered. This can be attributed to the charm PDF being treated in a different manner in the NNPDF3.1 fit as in the two other parton density fits that we consider.

For configurations in which the $t$-channel contributions play a role, we observe that the theoretical uncertainties are different from the NLO-QCD case. This comes from different dominant partonic initial states and a different scale dependence at the matrix-element level. In this way, for scenario $a_1$ NLO w/ $t$-channel predictions receive slightly smaller scale uncertainties and much larger PDF uncertainties. The origin of this behaviour is twofold. First, QCD-induced leptoquark pair production is dominated by gluon fusion topologies while leptoquark pair production via $t$-channel lepton exchanges is driven by charm-anticharm scattering. Second, the QCD contributions already depend on the strong coupling $g_s$ at LO, while the $t$-channel lepton exchange diagrams are independent of $g_s$.

The ratio $K_\text{NNLL}$ defined in eq.~\eqref{eq:Knnll} encompasses the full set of corrections under consideration, {\it i.e.}\ those originating from the $t$-channel diagrams, soft-gluon resummation and the use of NNLO parton densities. We find that for the considered scenarios and the CT18 and MSHT20 parton densities, the additional corrections lead to a further increase of the cross section. The $K_\text{NNLL}$ value indeed lies between 1.49 and 1.59 for scenario $a_1$ and between 1.13 and 1.17 for scenario $a_2$. The NNLL effects are thus as large as the $t$-channel ones for the first scenario, and the full corrections are purely NNLL-related for the second scenario. The NNPDF3.1 parton densities are extracted following a different treatment of the charm as for the CT18 and MSHT20 densities, thus different predictions could be expected. This is indeed what we obtain. In the case of scenario $a_1$, NNLL corrections lead to a reduction of the cross section, with $K_\text{NNLL}=1.60 < K_{\rm wt}=1.70$. The charm density contributing only in a subleading manner for scenario $a_2$, we then get more similar values of $K_\text{NNLL}=1.05 \sim K_{\rm wt} =1.01$.

The resummation of the soft-gluon corrections leads to a substantial reduction of the scale uncertainties relative to the NLO w/ $t$-channel results, as expected by virtue of the inclusion of higher-order contributions through their exponentiation. In contrast, PDF uncertainties are most of the time larger at NLO+NNLL than at NLO, as our NLO w/ $t$-channel+NNLL cross sections are obtained with NNLO PDF that are associated with different PDF errors. This last source of uncertainties constitutes the bulk of the theory error. However, one should keep in mind that more LHC data will be analysed in the future, opening the door to a substantial improvement of the PDF fits.

Overall the $t$-channel contributions are very important for the class of $R_2$ leptoquark scenarios favoured by the flavour anomalies. They can possibly lead to sizeable corrections to leptoquark pair-production cross sections, and should therefore be estimated appropriately so that their relevance could be verified. In contrast, soft-gluon resummation corrections are always significant and should be accounted for in any precision calculation. They impact both the rates themselves, but also lead to a significant reduction of the associated scale uncertainties.

\renewcommand{\arraystretch}{1.5}\setlength\tabcolsep{5pt}
\begin{table}
 \centering
 \resizebox{\textwidth}{!}{
 \begin{tabular}{l | l | l | lll | ll}
  & LQ & PDF & QCD (fb) & NLOwt (fb) & NNLLwt (fb) & $K_\text{wt}$ & $K_\text{NNLL}$
  \tabularnewline
  \hline
  \multirow{6}{*}{$b_1$} & \multirow{3}{*}{$R_2$} & CT18 & $0.700^{+12.0\%}_{-14.0\%}{}^{+14.8\%}_{-14.8\%}$ & $0.705^{+11.7\%}_{-14.0\%}{}^{+15.0\%}_{-15.0\%}$ & $0.846^{+7.0\%}_{-5.8\%}{}^{+21.0\%}_{-21.0\%}$ & $1.01$ & $1.21$
  \tabularnewline
  & & NNPDF3.1 & $0.641^{+12.1\%}_{-14.1\%}{}^{+7.6\%}_{-7.6\%}$ & $0.664^{+11.4\%}_{-13.5\%}{}^{+8.2\%}_{-8.2\%}$ & $0.683^{+7.8\%}_{-5.6\%}{}^{+8.9\%}_{-8.9\%}$ & $1.04$ & $1.07$
  \tabularnewline
  & & MSHT20 & $0.707^{+11.9\%}_{-14.0\%}{}^{+6.3\%}_{-6.3\%}$ & $0.713^{+11.7\%}_{-13.8\%}{}^{+6.4\%}_{-6.4\%}$ & $0.835^{+7.0\%}_{-5.7\%}{}^{+7.2\%}_{-7.2\%}$ & $1.01$ & $1.18$
  \tabularnewline
  \cline{2-8}
  & \multirow{3}{*}{$S_3$} & CT18 & $0.0108^{+14.1\%}_{-15.7\%}{}^{+25.2\%}_{-25.2\%}$ & $0.0108^{+14.0\%}_{-15.6\%}{}^{+25.2\%}_{-25.2\%}$ & $0.0145^{+8.5\%}_{-5.7\%}{}^{+40.8\%}_{-40.8\%}$ & $1.00$ & $1.33$
  \tabularnewline
  & & NNPDF3.1 & $0.0076^{+15.9\%}_{-17.1\%}{}^{+57.4\%}_{-57.4\%}$ & $0.0076^{+16.1\%}_{-17.0\%}{}^{+57.3\%}_{-57.3\%}$ & $0.0078^{+13.1\%}_{-8.9\%}{}^{+83.0\%}_{-83.0\%}$ & $1.00$ & $1.04$
  \tabularnewline
  & & MSHT20 & $0.0109^{+14.1\%}_{-15.8\%}{}^{+11.8\%}_{-11.8\%}$ & $0.0110^{+13.5\%}_{-16.0\%}{}^{+12.0\%}_{-12.0\%}$ & $0.0149^{+7.9\%}_{-5.6\%}{}^{+13.3\%}_{-13.3\%}$ & $1.01$ & $1.37$
  \tabularnewline
  \hline
  \multirow{12}{*}{$b_2$} & \multirow{3}{*}{$R_2$} & CT18 & $0.700^{+12.0\%}_{-14.0\%}{}^{+14.8\%}_{-14.8\%}$ & $0.720^{+11.5\%}_{-13.5\%}{}^{+15.6\%}_{-15.6\%}$ & $0.877^{+6.6\%}_{-5.6\%}{}^{+22.5\%}_{-22.5\%}$ & $1.03$ & $1.25$
  \tabularnewline
  & & NNPDF3.1 & $0.641^{+12.1\%}_{-14.1\%}{}^{+7.6\%}_{-7.6\%}$ & $0.733^{+9.9\%}_{-11.7\%}{}^{+12.9\%}_{-12.9\%}$ & $0.730^{+7.3\%}_{-4.9\%}{}^{+13.5\%}_{-13.5\%}$ & $1.14$ & $1.14$
  \tabularnewline
  & & MSHT20 & $0.707^{+11.9\%}_{-14.0\%}{}^{+6.3\%}_{-6.3\%}$ & $0.728^{+11.1\%}_{-13.6\%}{}^{+6.7\%}_{-6.7\%}$ & $0.862^{+6.7\%}_{-5.5\%}{}^{+7.7\%}_{-7.7\%}$ & $1.03$ & $1.22$
  \tabularnewline
  \cline{2-8}
  & \multirow{3}{*}{$S_3^{(+2/3)}$} & CT18 & $0.0108^{+14.1\%}_{-15.7\%}{}^{+25.2\%}_{-25.2\%}$ & $0.0108^{+13.9\%}_{-15.6\%}{}^{+25.2\%}_{-25.2\%}$ & $0.0145^{+8.2\%}_{-5.8\%}{}^{+40.8\%}_{-40.8\%}$ & $1.00$ & $1.34$
  \tabularnewline
  & & NNPDF3.1 & $0.0076^{+15.9\%}_{-17.1\%}{}^{+57.4\%}_{-57.4\%}$ & $0.0076^{+15.7\%}_{-17.1\%}{}^{+57.3\%}_{-57.3\%}$ & $0.0078^{+13.2\%}_{-8.8\%}{}^{+83.0\%}_{-83.0\%}$ & $1.00$ & $1.02$
  \tabularnewline
  & & MSHT20 & $0.0109^{+14.1\%}_{-15.8\%}{}^{+11.8\%}_{-11.8\%}$ & $0.0109^{+14.0\%}_{-15.9\%}{}^{+11.8\%}_{-11.8\%}$ & $0.0148^{+8.0\%}_{-5.7\%}{}^{+13.1\%}_{-13.1\%}$ & $1.00$ & $1.36$
  \tabularnewline
  \cline{2-8}
  & \multirow{3}{*}{$S_3^{(-1/3)}$} & CT18 & $0.0108^{+14.1\%}_{-15.7\%}{}^{+25.2\%}_{-25.2\%}$ & $0.0109^{+13.5\%}_{-15.7\%}{}^{+25.4\%}_{-25.4\%}$ & $0.0146^{+8.3\%}_{-5.7\%}{}^{+41.4\%}_{-41.4\%}$ & $1.01$ & $1.35$
  \tabularnewline
  & & NNPDF3.1 & $0.0076^{+15.9\%}_{-17.1\%}{}^{+57.4\%}_{-57.4\%}$ & $0.0077^{+15.6\%}_{-17.0\%}{}^{+57.0\%}_{-57.0\%}$ & $0.0078^{+13.0\%}_{-8.8\%}{}^{+82.5\%}_{-82.5\%}$ & $1.00$ & $1.03$
  \tabularnewline
  & & MSHT20 & $0.0109^{+14.1\%}_{-15.8\%}{}^{+11.8\%}_{-11.8\%}$ & $0.0110^{+13.9\%}_{-15.8\%}{}^{+12.0\%}_{-12.0\%}$ & $0.0149^{+7.8\%}_{-5.6\%}{}^{+13.3\%}_{-13.3\%}$ & $1.00$ & $1.37$
  \tabularnewline
  \cline{2-8}
  & \multirow{3}{*}{$S_3^{(-4/3)}$} & CT18 & $0.0108^{+14.1\%}_{-15.7\%}{}^{+25.2\%}_{-25.2\%}$ & $0.0110^{+13.6\%}_{-15.2\%}{}^{+26.0\%}_{-26.0\%}$ & $0.0149^{+8.2\%}_{-5.4\%}{}^{+43.4\%}_{-43.4\%}$ & $1.02$ & $1.38$
  \tabularnewline
  & & NNPDF3.1 & $0.0076^{+15.9\%}_{-17.1\%}{}^{+57.4\%}_{-57.4\%}$ & $0.0077^{+15.5\%}_{-16.8\%}{}^{+56.9\%}_{-56.9\%}$ & $0.0080^{+12.9\%}_{-8.7\%}{}^{+80.9\%}_{-80.9\%}$ & $1.01$ & $1.04$
  \tabularnewline
  & & MSHT20 & $0.0109^{+14.1\%}_{-15.8\%}{}^{+11.8\%}_{-11.8\%}$ & $0.0111^{+13.9\%}_{-15.3\%}{}^{+12.3\%}_{-12.3\%}$ & $0.0153^{+7.6\%}_{-5.3\%}{}^{+13.9\%}_{-13.9\%}$ & $1.02$ & $1.41$
 \end{tabular}
 }
 \caption{Same as in table~\ref{tab:benchmarksR2xsections} but for the benchmark scenarios $b_1$ and $b_2$ of section~\ref{sec:benchmr2s3}. As the cross sections for the $R_2$ mass eigenstates only differ marginally, they are provided as single predictions. For the same reason all cross sections for the $S_3$ mass eigenstates are provided collectively for benchmark point $b_1$.}
 \label{tab:benchmarksR2plusS3xsections}
\end{table}

We now study cross section predictions associated with the benchmark scenarios $b_1$ and $b_2$ introduced in section~\ref{sec:benchmr2s3}. In this case, a given benchmark features a large set of $R_2$ and $S_3$ leptoquark eigenstates of different electric charge and with non-vanishing Yukawa couplings whose values are given in table~\ref{tab:benchmarksR2plusS3}. The pair-production cross sections related to all different mass eigenstates are shown in table~\ref{tab:benchmarksR2plusS3xsections}. For those cross sections that only differ marginally ({\it i.e.}\ those associated with all $R_2$ mass eigenstates of a given benchmark, and those associated with all $S_3$ mass eigenstates in scenario $b_1$), only one entry in the table is provided.

In the two considered scenarios $b_1$ and $b_2$, the relevant $S_3$ Yukawa couplings associated with second-generation quarks all take moderately small values, while those associated with the third-generation quarks are large. The $t$-channel contributions to the total cross section are thus expected to play a subleading role, as the corresponding diagrams are either suppressed by the smallness of the Yukawa couplings, or by that of the related parton densities. This is confirmed by our findings. The $t$-channel contributions provide at most about 1\% or 2\% correction for $S_3$ pair production. For the same reason, the theory uncertainties are similar in the NLO w/ $t$-channel and in the pure NLO-QCD cases. In contrast, the $R_2$ leptoquark exhibits much larger couplings to second-generation quarks, but those couplings associate different quark and lepton flavours and chiralities, as well as feature different complex phases. The $t$-channel lepton exchange diagrams are found to generally contribute to percent-level corrections, with the exception of predictions made with NNPDF3.1 parton densities. In this case, we find larger $K_{\rm wt}$ values reaching 1.04 for scenario $b_1$ and 1.14 for scenario $b_2$. The larger corrections again highlight the importance of the treatment of the charm density in the PDF fits. Similarly, the relevance of the $t$-channel contributions is strongly connected to the variations in the theory uncertainties observed when comparing NLO-QCD and NLO w/ $t$-channel predictions, for the reasons already mentioned above.

Since in the benchmark scenarios $b_1$ and $b_2$ the influence of the $t$-channel contributions on the total pair-production cross sections is rather small, the enhancement of the NLO w/ $t$-channel+NNLL cross sections with respect to the pure NLO-QCD one ({\it i.e.}\ the value of $K_\text{NNLL}$) is mainly due to soft-gluon resummation. The size of $K_\text{NNLL}$ lies between 1.02 and 1.41, with larger values obtained for the CT18 and MSHT20 PDF sets. Predictions with the NNPDF3.1 set suffer from suppression effects arising from the switch from NLO densities (used for the NLO w/ $t$-channel calculations) to the NNLO ones (used for the NLO w/ $t$-channel+NNLL rates), as discussed previously. Similarly to the observations made in the analysis of the predictions relevant for the benchmarks $a_1$ and $a_2$, our most precise predictions at the NLO w/ $t$-channel+NNLL level exhibit smaller scale uncertainties than those at the NLO w/ $t$-channel one. The PDF errors increase, sometimes significantly, when NNLO PDFs are used.

\renewcommand{\arraystretch}{1.5}\setlength\tabcolsep{5pt}
\begin{table}
\centering
 \resizebox{\textwidth}{!}{
 \begin{tabular}{l | l | l | lll | ll}
  & LQ & PDF & QCD (fb) & NLOwt (fb) & NNLLwt (fb) & $K_\text{wt}$ & $K_\text{NNLL}$
  \tabularnewline
  \hline
  \multirow{6}{*}{$c_1$} & \multirow{3}{*}{$S_1^{(-1/3)}$} & CT18 & $1.35^{+11.8\%}_{-13.8\%}{}_{-13.6\%}^{+13.6\%}$ & $1.46^{+10.4\%}_{-12.5\%}{}^{+14.9\%}_{-14.9\%}$ & $1.75^{+6.3\%}_{-4.8\%}{}^{+20.9\%}_{-20.9\%}$ & $1.08$ & $1.30$
  \tabularnewline
  & & NNPDF3.1 & $1.26^{+12.0\%}_{-14.0\%}{}_{-5.9\%}^{+5.9\%}$ & $1.57^{+8.3\%}_{-10.5\%}{}^{+12.5\%}_{-12.5\%}$ & $1.61^{+5.9\%}_{-3.9\%}{}^{+12.9\%}_{-12.9\%}$ & $1.25$ & $1.28$
  \tabularnewline
  & & MSHT20 & $1.36^{+11.7\%}_{-13.8\%}{}^{+5.7\%}_{-5.7\%}$ & $1.48^{+10.4\%}_{-12.5\%}{}^{+6.3\%}_{-6.3\%}$ & $1.73^{+6.2\%}_{-4.9\%}{}^{+7.3\%}_{-7.3\%}$ & $1.09$ & $1.27$
  \tabularnewline
  \cline{2-8}
  & \multirow{3}{*}{$S_3$} & CT18 & $1.35^{+11.8\%}_{-13.8\%}{}_{-13.6\%}^{+13.6\%}$ & $1.35^{+11.7\%}_{-13.9\%}{}^{+13.6\%}_{-13.6\%}$ & $1.59^{+6.9\%}_{-6.0\%}{}^{+18.6\%}_{-18.6\%}$ & $1.00$ & $1.18$
  \tabularnewline
  & & NNPDF3.1 & $1.26^{+12.0\%}_{-14.0\%}{}_{-5.9\%}^{+5.9\%}$ & $1.26^{+11.7\%}_{-14.2\%}{}^{+5.9\%}_{-5.9\%}$ & $1.30^{+7.6\%}_{-5.9\%}{}^{+6.1\%}_{-6.1\%}$ & $1.00$ & $1.03$
  \tabularnewline
  & & MSHT20 & $1.36^{+11.7\%}_{-13.8\%}{}^{+5.7\%}_{-5.7\%}$ & $1.37^{+11.7\%}_{-13.7\%}{}^{+5.7\%}_{-5.7\%}$ & $1.58^{+7.0\%}_{-5.8\%}{}^{+6.4\%}_{-6.4\%}$ & $1.01$ & $1.16$
  \tabularnewline
  \hline
  \multirow{12}{*}{$c_2$} & \multirow{3}{*}{$S_1^{(-1/3)}$} & CT18 & $1.35^{+11.8\%}_{-13.8\%}{}_{-13.6\%}^{+13.6\%}$ & $1.44^{+10.7\%}_{-12.5\%}{}^{+14.7\%}_{-14.7\%}$ & $1.73^{+6.4\%}_{-5.0\%}{}^{+20.8\%}_{-20.8\%}$ & $1.07$ & $1.28$
  \tabularnewline
  & & NNPDF3.1 & $1.26^{+12.0\%}_{-14.0\%}{}_{-5.9\%}^{+5.9\%}$ & $1.53^{+8.6\%}_{-10.8\%}{}^{+12.2\%}_{-12.2\%}$ & $1.57^{+6.0\%}_{-4.1\%}{}^{+12.5\%}_{-12.5\%}$ & $1.21$ & $1.25$
  \tabularnewline
  & & MSHT20 & $1.36^{+11.7\%}_{-13.8\%}{}^{+5.7\%}_{-5.7\%}$ & $1.46^{+10.6\%}_{-12.7\%}{}^{+6.2\%}_{-6.2\%}$ & $1.71^{+6.2\%}_{-4.9\%}{}^{+7.2\%}_{-7.2\%}$ & $1.07$ & $1.26$
  \tabularnewline
  \cline{2-8}
  & \multirow{3}{*}{$S_3^{(+2/3)}$} & CT18 & $1.35^{+11.8\%}_{-13.8\%}{}_{-13.6\%}^{+13.6\%}$ & $1.40^{+11.3\%}_{-12.8\%}{}^{+14.6\%}_{-14.6\%}$ & $1.70^{+6.4\%}_{-5.4\%}{}^{+20.8\%}_{-20.8\%}$ & $1.04$ & $1.26$
  \tabularnewline
  & & NNPDF3.1 & $1.26^{+12.0\%}_{-14.0\%}{}_{-5.9\%}^{+5.9\%}$ & $1.46^{+9.2\%}_{-11.4\%}{}^{+12.6\%}_{-12.6\%}$ & $1.45^{+7.0\%}_{-4.6\%}{}^{+13.0\%}_{-13.0\%}$ & $1.16$ & $1.15$
  \tabularnewline
  & & MSHT20 & $1.36^{+11.7\%}_{-13.8\%}{}^{+5.7\%}_{-5.7\%}$ & $1.41^{+11.2\%}_{-12.9\%}{}^{+6.2\%}_{-6.2\%}$ & $1.66^{+6.6\%}_{-5.4\%}{}^{+7.1\%}_{-7.1\%}$ & $1.04$ & $1.22$
  \tabularnewline
  \cline{2-8}
  & \multirow{3}{*}{$S_3^{(-1/3)}$} & CT18 & $1.35^{+11.8\%}_{-13.8\%}{}_{-13.6\%}^{+13.6\%}$ & $1.39^{+11.2\%}_{-13.3\%}{}^{+13.9\%}_{-13.9\%}$ & $1.65^{+6.6\%}_{-5.8\%}{}^{+19.3\%}_{-19.3\%}$ & $1.03$ & $1.22$
  \tabularnewline
  & & NNPDF3.1 & $1.26^{+12.0\%}_{-14.0\%}{}_{-5.9\%}^{+5.9\%}$ & $1.37^{+10.2\%}_{-12.6\%}{}^{+6.8\%}_{-6.8\%}$ & $1.43^{+6.9\%}_{-4.9\%}{}^{+7.1\%}_{-7.1\%}$ & $1.09$ & $1.13$
  \tabularnewline
  & & MSHT20 & $1.36^{+11.7\%}_{-13.8\%}{}^{+5.7\%}_{-5.7\%}$ & $1.41^{+11.2\%}_{-13.2\%}{}^{+5.9\%}_{-5.9\%}$ & $1.64^{+6.5\%}_{-5.6\%}{}^{+6.7\%}_{-6.7\%}$ & $1.04$ & $1.21$
  \tabularnewline
  \cline{2-8}
  & \multirow{3}{*}{$S_3^{(-4/3)}$} & CT18 & $1.35^{+11.8\%}_{-13.8\%}{}_{-13.6\%}^{+13.6\%}$ & $1.49^{+9.9\%}_{-12.4\%}{}^{+16.2\%}_{-16.2\%}$ & $1.75^{+6.3\%}_{-4.9\%}{}^{+21.0\%}_{-21.0\%}$ & $1.10$ & $1.30$
  \tabularnewline
  & & NNPDF3.1 & $1.26^{+12.0\%}_{-14.0\%}{}_{-5.9\%}^{+5.9\%}$ & $1.57^{+8.3\%}_{-10.9\%}{}^{+11.7\%}_{-11.7\%}$ & $1.71^{+5.3\%}_{-3.3\%}{}^{+12.8\%}_{-12.8\%}$ & $1.25$ & $1.36$
  \tabularnewline
  & & MSHT20 & $1.36^{+11.7\%}_{-13.8\%}{}^{+5.7\%}_{-5.7\%}$ & $1.52^{+10.1\%}_{-12.1\%}{}^{+6.4\%}_{-6.4\%}$ & $1.77^{+5.9\%}_{-4.7\%}{}^{+7.4\%}_{-7.4\%}$ & $1.12$ & $1.30$
 \end{tabular}
 }
 \caption{Same as in table~\ref{tab:benchmarksR2xsections} but for the benchmark scenarios $c_1$ and $c_2$ of section~\ref{sec:benchms1s3}. For benchmark point $c_1$ the cross sections associated with the production of any pair of $S_3$ mass eigenstates only differ marginally, and are thus displayed collectively.}
 \label{tab:benchmarksS1plusS3xsections}
\end{table}

The final benchmark scenarios that we discuss in this subsection are the $c_1$ and $c_2$ scenarios introduced in section~\ref{sec:benchms1s3}, featuring both the $S_1$ and $S_3$ leptoquarks. We consider two setups for which all non-zero values of the leptoquark Yukawa couplings are given in table~\ref{tab:benchmarksS1plusS3}. In table~\ref{tab:benchmarksS1plusS3xsections} we present the associated pair-production cross sections for both benchmark points. Once again, we have grouped together individual cross sections that only marginally differ, namely all $S_3$ pair-production cross sections related to scenario $c_1$.

In the $c_1$ scenario we obtain values of $K_{\rm wt}$ for $S_1$ pair production that span the $[1.08, 1.25]$ range. This contrasts with the case of $S_3$ pair production, for which the $t$-channel diagrams are negligible. These properties can be explained by the coupling values (see table~\ref{tab:benchmarksS1plusS3}), as the $S_3$ state only weakly couples to quarks whereas the $S_1$ state dominantly couples to second-generation quarks with strengths given by $y_{1,23}^{LL}$ and $y_{1,23}^{RR}$. As an expected consequence, predictions relying on NNPDF3.1 densities are significantly different from those obtained with the CT18 or MSHT20 densities, with the $K_{\rm wt}$ value for the $pp\to S_1 S_1$ process much larger in the NNPDF3.1 case. For the second scenario $c_2$, both leptoquarks couple more strongly to second-generation quarks due to the large $y_{3,23}^{LL}$ value. The $t$-channel diagram contributions to the cross section are therefore expected to play a more important role for all production processes as demonstrated by the results shown in the table. Moreover, as $S_3$ eigenstates of different electric charge couple to strange quarks, charm quarks or both, the corresponding cross sections are found to differ by up to 15\%.

Depending on the PDF and the final state considered, the combined corrections due to the $t$-channel contributions and soft-gluon resummation (depicted by $K_\text{NNLL}$) vary between 1.03 and 1.36. While soft-gluon resummation contributions generally lead to an increase of the cross sections of approximately 20\% with respect to $K_\text{wt}$ for CT18 and MSHT20 PDFs, the impact is milder for the NNPDF3.1 predictions. This effect is related to the above-mentioned differences between the NLO and NNLO NNPDF3.1 fits. It mainly manifests itself in final states involving leptoquarks that couple to the charm quark. As the $S_3^{(-4/3)}$ leptoquark only couples to strange quarks, the corresponding $K_\text{NNLL}$ value is much larger than the values of $K_\text{NNLL}$ associated with the $S_1$ and $S_3^{(-1/3)}$ states that couple to both strange and charm quarks. Finally, we obtain a $K_\text{NNLL}$ value for $S_3^{(+2/3)}$ pair production that is smaller than 1. This is a consequence of $S_3^{(+2/3)}$ solely coupling to the charm quark. The behaviour of the scale and PDF uncertainties is similar to what we have already observed for the other scenarios considered in this section. A more detailed discussion on the uncertainties is carried out in the next section.

\subsection{Impact on experimental limits: theoretical errors and extra contributions to rates}\label{sec:exp_impact}

\begin{figure}
\centering
\includegraphics[width=.5\textwidth]{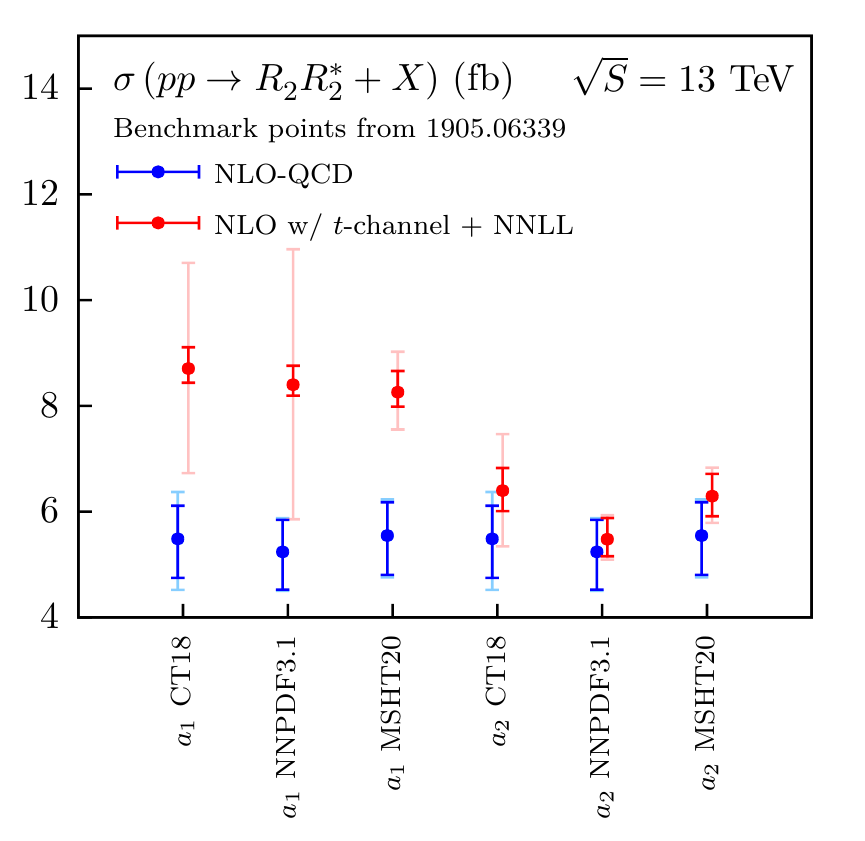}\includegraphics[width=.5\textwidth]{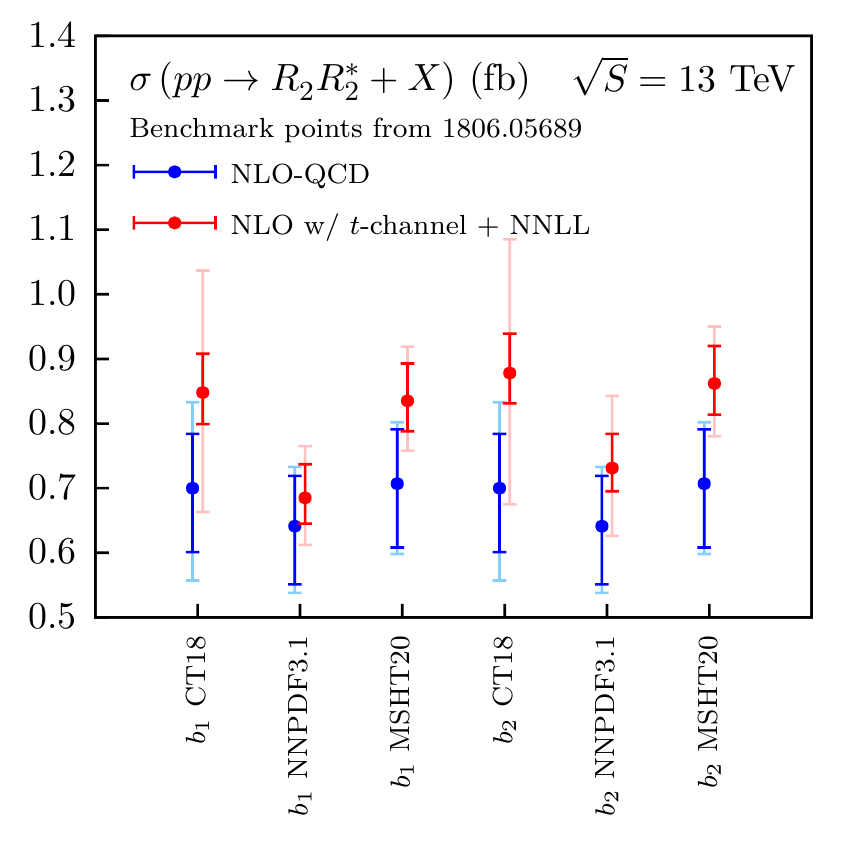}\\
\includegraphics[width=.5\textwidth]{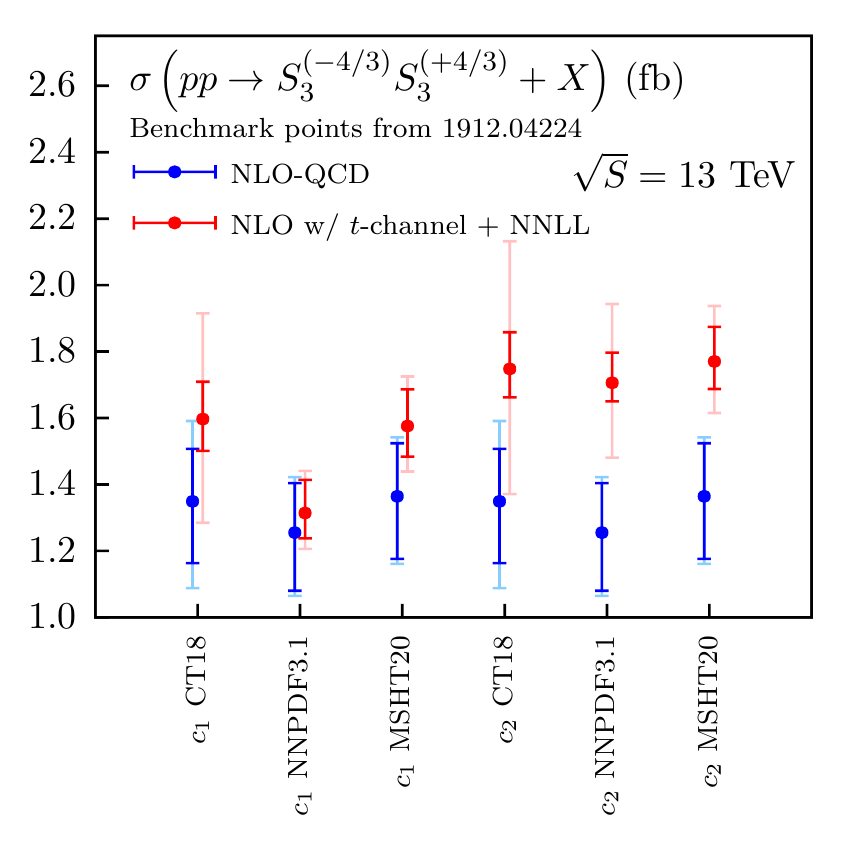}
\caption{Comparison of total cross section predictions at the NLO-QCD (blue) and NLO w/ $t$-channel + NNLL (red) accuracy, for the benchmark points discussed in tables \ref{tab:benchmarksR2xsections}, \ref{tab:benchmarksR2plusS3xsections}, and \ref{tab:benchmarksS1plusS3xsections}. The dark-coloured error bars denote the scale uncertainties, while the light-coloured ones combine them with the PDF uncertainties.}
\label{fig:benchmarksplots}
\end{figure}
In the previous section we have demonstrated that in certain models and for certain choices of benchmark points and PDF sets, $t$-channel contributions and soft-gluon resummation corrections can substantially alter cross section predictions and lead to a reduction of the scale uncertainties. In contrast, the PDF uncertainties are found either similar to or larger than those inherent to the fixed-order results. To emphasise these findings more clearly, we collect the results of tables~\ref{tab:benchmarksR2xsections}, \ref{tab:benchmarksR2plusS3xsections} and \ref{tab:benchmarksS1plusS3xsections} and present them graphically in figure \ref{fig:benchmarksplots} for selected leptoquark production processes. We separately indicate the associated scale uncertainties and the full theoretical errors, that additionally contain PDF uncertainties.

All presented cases visually confirm the findings of section~\ref{sec:results:benchmarks}. Scale uncertainties decrease when NLO+NNLL corrections are accounted for, whereas the overall theory errors generally increase by virtue of larger uncertainties originating from the use of NNLO PDFs (employed for the NLO w/ $t$-channel + NNLL predictions) as compared with the NLO PDF sets (used for fixed-order predictions). Notably, the NLO w/ $t$-channel + NNLL predictions obtained with the most recent MSHT20 PDFs have the overall theory errors smaller than the NLO-QCD predictions. This originates from the NLO and NNLO PDF uncertainties being very similar, and demonstrates the potential of our calculations in significantly reducing the overall theory error once PDF uncertainties are under control.

\begin{figure}
\centering
\includegraphics[width=\textwidth]{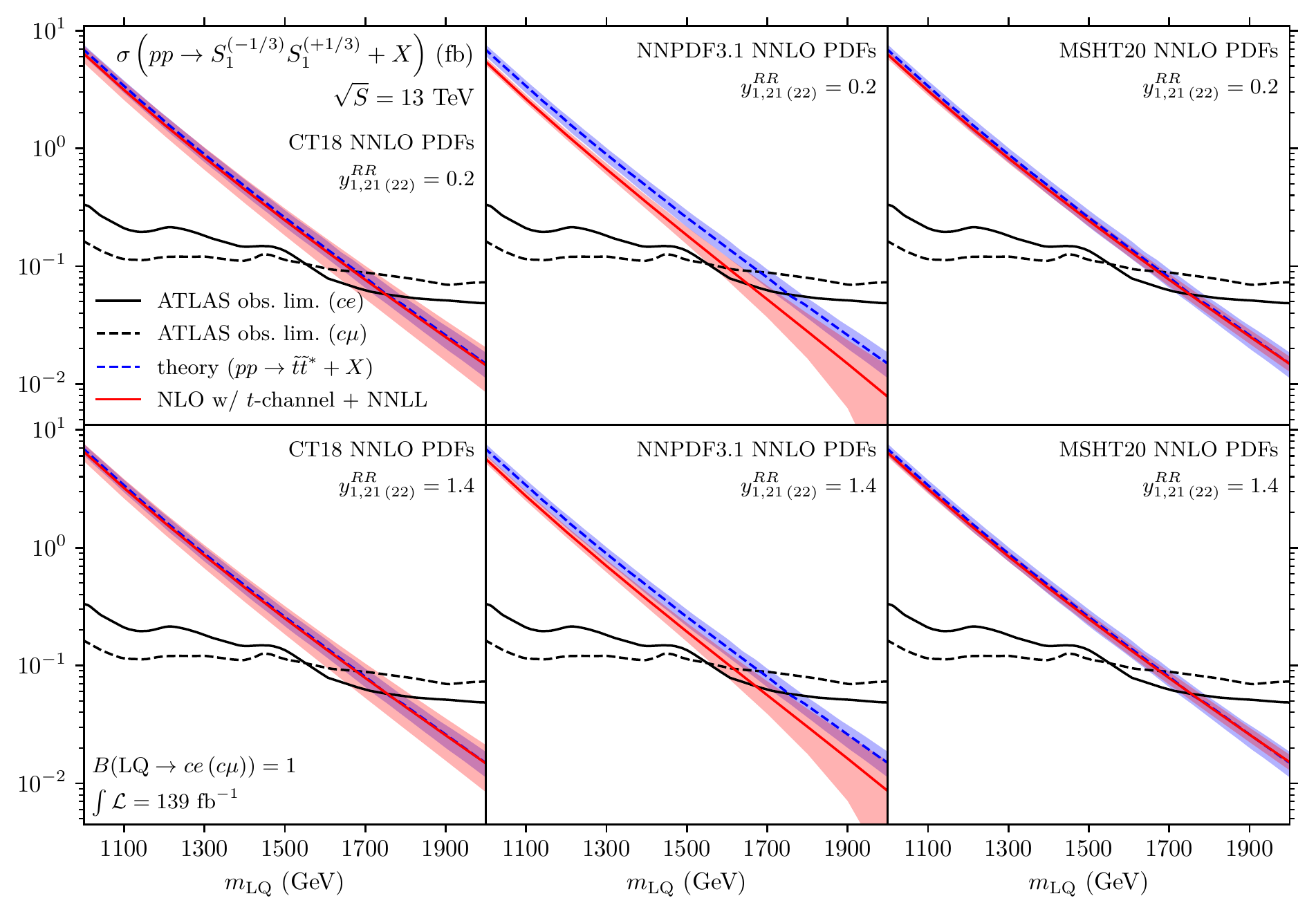}
\caption{Comparison of observed 95\% confidence level limits \cite{Aad:2020iuy} for leptoquark pair production when assuming a branching ratio of 1 for either the $\text{LQ} \to ce$ decay (solid black) or the $\text{LQ} \to c\mu$ decay (dashed black). We present the theory predictions used in the ATLAS analysis \cite{Aad:2020iuy} (dashed blue) and our NLO w/ $t$-channel + NNLL ones (solid red) calculated with the CT18 (left), NNPDF3.1 (middle), and MSHT20 (right) parton densities. Band thickness denotes combined PDF and scale uncertainties. For the NLO w/ $t$-channel + NNLL predictions, we have adopted leptoquark-charm-lepton coupling values of $y^{RR}_{1,21\,(22)} = 0.2$ (top row) and $y^{RR}_{1,21\,(22)} = 1.4$ (bottom row).}
\label{fig:comparison-ATLAS}
\end{figure}

Figure~\ref{fig:benchmarksplots} also clearly shows that there are models and benchmark points for which predictions including $t$-channel contributions as well as NNLL corrections do not agree with the NLO-QCD ones (within their respective theory errors). This calls for a particular caution when using theoretical predictions in experimental analyses to extract limits on models, as it is critical to make sure that these effects are accounted for. Not all predictions, however, are affected to the same extent. As discussed in the previous section there are many factors impacting the behaviour of the predicted cross sections, such as the studied final state, the values of the leptoquark Yukawa couplings and masses ({\it i.e.}\ the particular benchmark scenario) or the parton density functions that have been used for the numerical evaluation of the cross sections. Their complicated interplay makes it difficult to predict the overall effect on the cross sections. The decision if one should use NLO w/ $t$-channel + NNLL predictions or if it is reasonable to rely only on NLO-QCD cross sections has therefore to be made individually for each considered scenario.

In this context, it is important to investigate to what extent the mass exclusion limits obtained as a result of an experimental analysis could be affected by the corrections calculated in this work. While analysing multiple possible scenarios is beyond the scope of this paper, in figure \ref{fig:comparison-ATLAS} we show an example study in the context of a simple model featuring a single leptoquark species $(S_1^{-1/3})$. We enforce that the leptoquark either couples to a charm quark and a muon only, or couples cross-generationally to a charm quark and an electron, {\it i.e.}\ with only $y_{1,22}^{RR}\neq 0$ or only $y_{1,21}^{RR}\neq 0$ respectively. The results are presented for small Yukawa coupling strengths ($y_{1,2k}^{RR} = 0.2$; top row) and larger Yukawa couplings ($y_{1,2k}^{RR} = 1.4$; bottom row). Even in this simple model, the exclusion limits of~\cite{Aad:2020iuy} can be lowered by more than 50 GeV. The size of the shift depends to a large degree on the PDF set that is used. In this particular case the effect driven by the choice of using the NNPDF3.1 set can be traced back to the corresponding treatment of the charm quark distribution. Moreover, as already pointed out, predictions computed with the most recent parton density fits ({\it i.e.}\ MSHT20) lead to a smaller overall theory error. It is however clear that admitting more non-zero couplings could easily lead to bigger effects (as shown by the results of sections~\ref{sec:results:individualimpact} and \ref{sec:results:benchmarks}), as well as have an impact on exclusion limits for third-generation leptoquarks. We leave the corresponding detailed studies for future work.

\section{Conclusions}\label{sec:conclusions}
In this work, we have performed a comprehensive study of the currently most precise theoretical predictions for scalar leptoquark pair production at the LHC. As reported here and in an earlier letter~\cite{Borschensky:2020hot}, these predictions contain, apart from pure QCD contributions, contributions of diagrams featuring a $t$-channel lepton exchange and their interference with the QCD diagrams. While all the components of the fixed-order results have been evaluated at NLO in QCD, we have additionally included corrections due to soft-gluon resummation up to NNLL accuracy. The classes of the corrections considered here become particularly important for leptoquark models with large masses and large Yukawa couplings. Since such leptoquark scenarios can account for the measured lepton flavour anomalies and the long-standing discrepancies in the anomalous magnetic moment of the muon, the aforementioned models have been enjoying a lot of interests lately. In this spirit, we have provided and carefully analysed cross section predictions in the framework of selected benchmark scenarios recently proposed in the literature.

Our analysis has shown that the considered corrections are important and often induce effects of tens of percents. In the models considered here, corrections up to 60\% are observed. Notably, we have found that the theoretical errors of pure NLO-QCD results (without soft-gluon resummation and the inclusion of the $t$-channel lepton exchange diagrams) often do not account for the size of the $t$-channel and the NNLL corrections. We therefore recommend to check the size of these contributions before using NLO-QCD predictions alone in any analysis. We have also demonstrated that the final size of the (sum of all) corrections originates from a very delicate interplay between different factors. These include the type of leptoquarks considered, their masses, the flavour pattern of their Yukawa couplings, as well as the parton distribution functions used to obtain the numerical predictions. This interplay manifests itself through various distinct higher-order effects often coming with opposite signs and different sizes, be it the $t$-channel contributions, the soft-gluon corrections or the choice between an NLO or an NNLO parton density set. Consequently, it is difficult to predict, on a general basis, the overall modification expected for total rate predictions, making it necessary to perform dedicated calculations for each individual scenario that one may want to consider.

In addition to potentially modifying the cross section in a sizable manner, the calculated corrections also lead to a substantial reduction of the scale dependence of the predictions. Although for higher leptoquark masses the theoretical error is in most cases still dominated by the NNLO PDF error, the situation is expected to change with newer releases of NNLO parton densities. It will eventually resemble the one arising for pure NLO-QCD predictions in which the PDF error covers only a small portion of the full theory uncertainties. In fact, such a situation is already observed for the latest MSHT20 NNLO set. Therefore one can certainly expect that the relative gain in precision due to the calculations described in this report will only grow with time.

The calculation of the NLO $t$-channel contributions presented in this paper have been performed with the help of two public codes, \mgamc\ and the \pwb\ framework, that we have modified according to the guidelines shown in section~\ref{sec:numerics}. To calculate NNLL soft-gluon corrections, a dedicated code is provided. Both this code and the modified versions of \mgamc\ and \pwb\ can be downloaded from the webpage
\url{https://www.uni-muenster.de/Physik.TP/research/kulesza/leptoquarks.html}.

\section*{Acknowledgements}
We thank Julien Baglio for discussions regarding the modifications of the \pwb{}, as well as Valentin Hirschi, Olivier Mattelear, Hua-Sheng Shao and Marco Zaro for their help with mixed-order computations in \mgamc. We are moreover grateful to Olcyr Sumensari and Nejc Ko\v{s}nik for providing information about the fit that has been performed in ref.~\cite{Becirevic:2018afm} and for their private update~\cite{Becirevic:2020pc} in the light of the newest measurements, as well as to Richard Ruiz for enlightening discussions over the course of this project (relative in particular to the reduced PDF errors associated with the MSHT20 set of parton densities). We furthermore acknowledge support by the state of Baden-W\"urttemberg through bwHPC and the DFG Grant No.\ INST 39/963-1 FUGG (bwForCluster NEMO). The work of DS was supported in part by the DFG Grant No. KU3103/2. Numerical calculations have been also performed on the PALMA HPC cluster of the WWU M\"unster.

\appendix

\section{\mgamc{} implementation}\label{app:mg}
In order to use \mgamc{} to evaluate the virtual contributions with its {\sc MadLoop} module~\cite{Hirschi:2011pa} and combine the result with the real emission contributions following the FKS subtraction method~\cite{Frixione:1995ms} as embedded in {\sc MadFKS}~\cite{Frederix:2009yq}, we need to modify two of the core files of the program. In the file  \texttt{base\_objects.py} we add two lines in the function \texttt{is\_perturbating}, right at the beginning of the first \texttt{for}-loop,
\begin{verbatim}
for int in model['interactions'].get_type('base'):
  if len(int.get('orders'))>1:
    continue

  ## BEGIN ADDITION
  if order in int.get('orders').keys() and \
    abs(self.get('pdg_code')) in [11,12,13,14,15,16]:
      return True
  ## END ADDITION

  if order in int.get('orders').keys() and self.get('pdg_code') in \
    [part.get('pdg_code') for part in int.get('particles')]:
      return True
\end{verbatim}

In the file \texttt{loop\_diagram\_generation.py} we modify the \texttt{user\_filter} method at two different places. First, we set the \texttt{edit\_filter\_manually} variable to \texttt{True}. Second, we add, in the loop over the virtual diagrams, the following lines,
\begin{verbatim}
for diag in self['loop_diagrams']:

  [...]

  # Apply the custom filter specified if any
  if filter_func:
    try:
      valid_diag = filter_func(diag, structs, model, i)
    except Exception as e:
      raise InvalidCmd("The user-defined filter '%s' did not"%filter+
        " returned the following error:\n       > %s"%str(e))

  ## BEGIN ADDITION
  loop_pdgs = [abs(x) for x in diag.get_loop_lines_pdgs()]
  is_loop_lepton = (len([x for x in loop_pdgs if x in [11,12,13,14,15,16]])>0)
  is_loop_gluon  = (21 in loop_pdgs)
  if is_loop_lepton and not is_loop_gluon:
    valid_diag=False

  connected_id = diag.get_pdgs_attached_to_loop(structs)
  isnot_lepton_correction = (len([x for x in connected_id if not abs(x) in \
    [11,12,13,14,15,16] ])>0)
  if not isnot_lepton_correction:
    valid_diag=False
  ## END ADDITION
\end{verbatim}

For some specific leptoquark benchmark scenarios (not considered in this paper), it is possible that \mgamc{} is unable to produce a numerical code suitable for NLO calculations. This is connected to a bug in the treatment of the fermion flow in the {\tt helas\_objects.py} file. This bug can be fixed by modifying the {\tt check\_majorana\_and\_flip\_flow} method, in which the line
\begin{verbatim}
  new_wf = wavefunctions[wavefunctions.index(new_wf)]
\end{verbatim}
must be replaced by
\begin{verbatim}
   if (not new_wf.get('is_loop')) or (new_wf.get('pdg_code')>0):
       index_wf = wavefunctions.index(new_wf)
   else:
       for i_wf, wf in enumerate(wavefunctions):
           if new_wf == wf and wf.get('pdg_code')==new_wf.get('pdg_code'):
               index_wf = i_wf
               break
       else:
           raise ValueError
   new_wf = wavefunctions[index_wf]
\end{verbatim}
This bug has been acknowledged by the \mgamc\ authors and will be fixed in future releases of the code.

\section{\pwb{} implementation}\label{app:pwb}

As only the particle identifiers of the Standard Model particles are hard-coded into the \pwb{}, several routines require modifications to account for the new electrically and colour charged leptoquarks included in our model. In our \pwb{} implementation, we choose as their particle identifiers codes between 770 and 780. We first append the leptoquarks to the list of charged particles, for which we modify the function \texttt{is\_charged} in the file \texttt{find\_regions.f}:
\begin{verbatim}
if(fl.eq.0) then

   [...]

!! BEGIN ADDITION
elseif (abs(fl).ge.770.and.abs(fl).le.780) then
   is_charged=.true.
!! END ADDITION

else
   is_charged=.false.
endif
\end{verbatim}
In the same file, we add the leptoquarks to the list of coloured particles in the function \texttt{is\_coloured}:
\begin{verbatim}
if(abs(fl).le.6) then

   [...]

!! BEGIN ADDITION
elseif (abs(fl).ge.770.and.abs(fl).le.780) then
   is_coloured=.true.
!! END ADDITION

else
   is_coloured=.false.
endif
\end{verbatim}
Furthermore, we need to adapt the soft-virtual term which is calculated in the file \texttt{sigsoftvirt.f} and the subroutine \texttt{btildevirt}. Thus, as it receives contributions from coloured massive final-state particles, we include the corresponding contribution for the leptoquarks in the loop over the final states:
\begin{verbatim}
do leg=3,nlegborn

   [...]

   if(is_coloured(fl)) then
      if(kn_masses(leg).eq.0) then

         [...]

      else

         !! BEGIN MODIFICATION
         if (abs(flst_born(leg,jb)).ge.770.and.abs(flst_born(leg,jb)).le.780) then
           Q = Q - cf*(ll-0.5*Intm_ep(pborn(0,leg)))
         else
           Q = Q - c(flst_born(leg,jb))*(ll-0.5*Intm_ep(pborn(0,leg)))
         endif
         !! END MODIFICATION

      endif
   endif
enddo
\end{verbatim}
With this modification, we make sure that correct colour coefficients are used for the leptoquarks (which, in our case, is the Casimir invariant $C_F$ of the fundamental representation of $SU(3)$).

\bibliographystyle{JHEP}
\bibliography{scalar-leptoquarks}

\end{document}